\documentclass[12pt]{article}

\usepackage{times}
\usepackage{fixltx2e}
\usepackage{graphicx}
\usepackage[table]{xcolor}
\usepackage{ragged2e}
\usepackage{multirow}
\usepackage{helvet}
\usepackage{booktabs}
\usepackage{colortbl}
\RequirePackage{mathpazo}
\usepackage{amsmath}

\usepackage{blindtext}

\usepackage[colorlinks = true,
            linkcolor = magenta,
            urlcolor  = magenta,
            citecolor = magenta,
            anchorcolor = magenta]{hyperref}

\makeatletter
\newenvironment{figurehere}
{\def\@captype{figure}}
{}
\makeatletter

\usepackage{ccaption}
\usepackage{siunitx}
\usepackage{fixltx2e}
\newenvironment{sciabstract}{%
\begin{quote} \bf}
{\end{quote}}

\newcommand{\cb}{\color{blue}}

\newcounter{lastnote}

\title{A Ferroelectric Compute-in-Memory Annealer for Combinatorial Optimization Problems}

\author
{Xunzhao Yin$^{1*}$, Yu Qian$^{1*}$,  Alptekin Vardar$^{2}$, Marcel G{\"u}nther$^{2}$, \\
Franz M{\"u}ller$^{2}$, Nellie Laleni$^{2}$, 
Zijian Zhao$^{3}$, Zhouhang Jiang$^{3}$, Zhiguo Shi$^{1}$, \\
Yiyu Shi$^{3}$,
Xiao Gong$^{4}$, Cheng Zhuo$^{1\dagger}$, 
Thomas K{\"a}mpfe$^{2\dagger}$, Kai Ni$^{3\dagger}$
\\
\normalsize{$^{1}$Zhejiang University, Hangzhou, China;} \\
\normalsize{$^{2}$Fraunhofer IPMS, Dresden, Germany;}\\
\normalsize{$^{3}$University of Notre Dame, Notre Dame, USA;}\\
\normalsize{$^{4}$National University of Singapore, Singapore;}
\\
\normalsize{$^\ast$Equal contributions;} \\
\normalsize{$^\dagger$To whom correspondence should be addressed; E-mail:}\\ \normalsize{czhuo@zju.edu.cn, thomas.kaempfe@ipms.fraunhofer.de, kni@nd.edu.}
}

\date{}

\begin{document} 

\maketitle 
\begin{sciabstract}
Computationally hard combinatorial optimization problems (COPs) are ubiquitous in many applications, including logistical planning,  resource allocation, chip design, drug explorations, and more.
Due to their critical significance and the inability of conventional hardware in efficiently handling scaled COPs, there is a growing interest in developing computing hardware tailored specifically for COPs, including digital annealers, dynamical Ising machines, and quantum/photonic systems. 
However, significant hurdles still remain, such as the memory access issue, the system scalability and restricted applicability to certain types of COPs, and VLSI-incompatibility, respectively.
Here, a ferroelectric field effect transistor (FeFET) based compute-in-memory (CiM) annealer is proposed. 
After converting COPs into quadratic unconstrained binary optimization (QUBO) formulations, a hardware-algorithm co-design is conducted, yielding an energy-efficient, versatile, and scalable hardware for COPs.
To accelerate the core vector-matrix-vector (VMV) multiplication of  QUBO formulations, 
a FeFET based CiM array is exploited, which can accelerate the intended operation in-situ due to its unique three-terminal structure.
In particular, a lossless compression technique is proposed to prune typically sparse QUBO matrix to reduce hardware cost. 
Furthermore, a multi-epoch simulated annealing (MESA) algorithm is proposed to replace conventional simulated annealing for its faster convergence and better solution quality. The effectiveness of the proposed techniques is validated through the utilization of developed chip prototypes for successfully solving graph coloring problem, indicating great promise of  FeFET CiM annealer in solving general COPs.

\end{sciabstract}

\section*{Introduction}
\label{sec:introduction}

Combinatorial optimization problems (COPs), as shown in Fig.\ref{fig:motivation}\textbf{a}, are prevalent in diverse fields, including logistics, resource allocation, communication network design, finance, drug discovery, and transportation systems, etc., \cite{yu2013industrial, paschos2014applications, naseri2020application, barahona1988application}.  
Often, these problems belong to the class of non-deterministic polynomial-time-hard (NP-hard) problems,  representing some of the most challenging computational tasks in  the NP domain.
Solving COPs using digital computers based on the von Neumann architecture poses difficulties, given the exponential growth in required resources regarding the computational power and latency as the problems scale up \cite{markov2014limits, markov2013know, greenlaw1995limits}. 
Therefore,  there is a pressing need to explore novel hardware design with alternative architectures and algorithms that can efficiently tackle COPs. 
This research frontier holds crucial implications for real-world applications, with the potential to address complex and resource-intensive problems with greater effectiveness.

\begin{figurehere}
	\centering
	\includegraphics [width=0.95\linewidth]{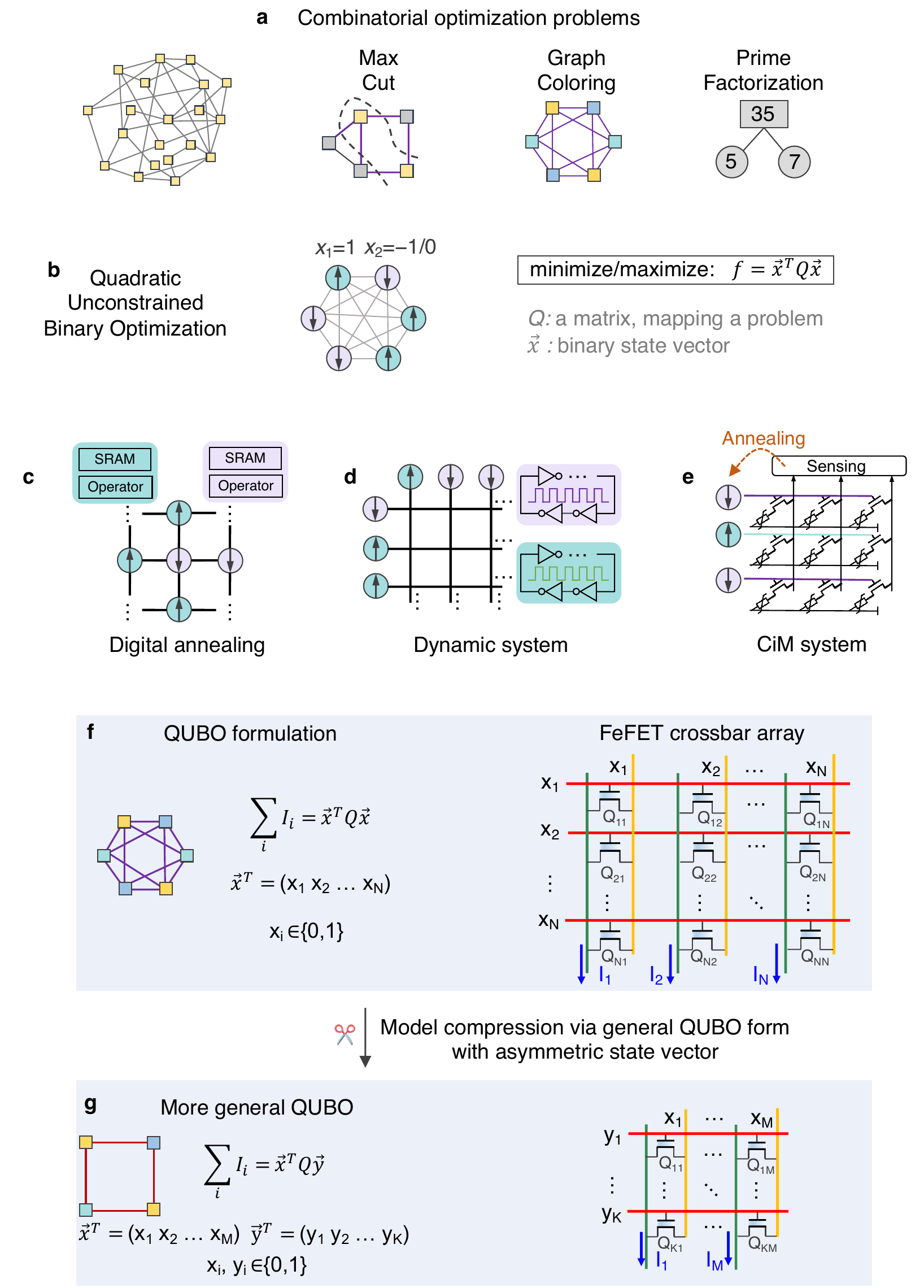}
	\caption{\textit{\textbf{Acceleration of solving COPs with CiM array.} \textbf{a.} COPs (e.g., Max-Cut, graph coloring, prime factorization etc.) can be converted to \textbf{b.} $\Vec{x}\ ^TQ\Vec{x}$
    QUBO formulation. Many hardware systems, including \textbf{c.} digital annealing system, \textbf{d.} dynamical system, and \textbf{e.} CiM system are promising for solving COPs. 
    \textbf{f.} Due to the unique characteristics of FeFET based CiM crossbar, it can implement typical $\Vec{x}\ ^TQ\Vec{x}$ QUBO formulation with symmetrical input vectors. \textbf{g.} The array can also accelerate a more general and compact $\Vec{x}\ ^TQ\Vec{y}$ QUBO formulation with asymmetrical vectors, thus achieving high efficiency and low cost.}}
	\label{fig:motivation}
\end{figurehere}

Many COPs, including graph coloring, Max-Cut, and traveling salesman problem, etc., can be mapped to Ising spin glass model, or often go by the name QUBO (i.e., quadratic unconstrained binary optimization) \cite{lucas2014ising}, which have emerged as a powerful framework for effectively modeling and solving a wide range of COPs \cite{mohseni2022ising}. 
In this framework, the problem variables are elegantly represented as spins, and the interactions or constraints between variables are represented as spin-to-spin couplings.
The objective function of the problem can then be mapped to the Hamiltonian energy function of the Ising model. 
The solution of a problem then corresponds to the combination of spins that minimizes the Ising Hamiltonian $H_{\mathrm{P}}$, which  can be formulated as follows:
\begin{equation}
\vspace{-1ex}
\label{equ:Ising model}
    \min H_{\mathrm{P}}=\sum_{i,j=1}^{N}J_{i j}\sigma_{i}\sigma_{j}+\sum_{i=1}^{N}h_{i}\sigma_{i}
\end{equation}
where $N$ denotes the number of spins, and $\sigma_i \in\{1, -1\}$ represents the state of spin $i$. $J_{ij}$ and $h_i$ stand for the coupling between spin $i$ and $j$ and the self-coupling of spin $i$, respectively. 
Through a simple variable change $\sigma_i = 1 - 2x_i$, $x_i \in \{0,1\}$, the Ising Hamiltonian (Eq. \ref{equ:Ising model}) can be readily transformed into a QUBO matrix form \cite{date2021qubo, zaman2021pyqubo} as  
\begin{equation}
\vspace{-1ex}
\label{equ:QUBO}
 H_{QUBO} = \Vec{x}\ ^TQ\Vec{x}
\end{equation}
where $\Vec{x} = (x_1, x_2, ... ,x_n)$, and $Q$ is a symmetric or an equivalent upper triangular $n \times n$ matrix \cite{glover2018tutorial,glover2022quantum}, as shown in Fig.\ref{fig:motivation}\textbf{b}. For instance, consider a Max-Cut problem defined on an undirected graph G(\textit{V},\textit{E}), where \textit{V} represents the set of vertices, and \textit{E} represents the set of edges \cite{glover2018tutorial}. 
The Max-Cut problem is mapped into the QUBO form by introducing a binary variable $x_i \in \{0,1\}$ for each vertex $\textit{i} \in V $, where $x_i$ takes the value 1 if vertex \textit{i} is assigned to one set and 0 if it belongs to the other set.
The objective function of the Max-Cut problem can then be formulated as follows:
\begin{equation}
\vspace{-1ex}
\label{equ:max cut1 intro}
\min \sum_{(i,j)\in E}(2x_i x_j - x_i - x_j)
\end{equation}
Considering the binary $x_i$, such form can be easily represented as the $\Vec{x}\ ^TQ\Vec{x}$ QUBO form \cite{glover2018tutorial,glover2022quantum}.
The conversion of other COPs, such as graph coloring problems and prime factorization problems, into the QUBO form 
is further elaborated in the Sec.\ref{sec:conversion} of Supplementary Information.

To solve COPs efficiently, various alternative computing hardware are under active research. Fig.\ref{fig:motivation}\textbf{c, d} briefly summarizes different electronic implementations. One class of hardware are digital ASIC annealers, where various annealing algorithms are implemented in digital circuits \cite{yamamoto20207, katsuki2022fast, onizawa2023local,takemoto20214}. Usually the spin coupling matrix is stored in memory and data need to be frequently transferred between memory and computing units for energy computation and annealing, which can be energy- and time-consuming as the problem scales up.
An attractive alternative is dynamical system Ising machines, where the intrinsic system dynamics and tendency to settle at lowest energy state is exploited to solve the COPs, as shown in Fig.\ref{fig:motivation}\textbf{d}. 
Once the spin coupling matrix are programmed within the hardware, these solvers naturally explore the solution space and ultimately find the  spin combination that minimizes the Ising energy without explicitly executing annealing algorithms.
Examples include the oscillator based Ising machine (OIM)  \cite{moy20221, ahmed2021probabilistic,dutta2021ising}, latch based Ising machine \cite{roychowdhury2021bistable, mallick2023cmos, afoakwa2021brim},  and optical based coherent ising machine (CoIM) \cite{pierangeli2019large, honjo2021100, yamamoto2020coherent, mcmahon2016fully, bohm2019poor}.
 
While the concept of such a system holds immense promise, there are several challenges remain to be addressed. 
First, the dynamics and robustness of dynamical Ising solvers is highly sensitive to the coupling implementations between spins, as a slight deviation in coupling strength can lead to convergence disruption of the solution \cite{ahmed2021probabilistic, mallick2023cmos}. Therefore, it poses a significant challenge in precisely mapping the spin coupling matrix into hardware. 
Second, exploiting dynamical Ising solvers to their full potential requires mapping the entire problem onto a single solver. For large scale problems that are beyond the capacity of the solver, how to efficiently map the problems to multiple separate chips and implement chip-to-chip communication while maintaining system dynamics require substantial work. Therefore, scaling of dynamical Ising solver is a critical challenge. Lastly, integration of self-interaction into these dynamical Ising solvers is not straightforward, thus allowing easy mapping of only a subclass of COPs without self-interaction terms, such as Max-Cut, Sherrington-Kirkpatrick models, etc \cite{moy20221, hamerly2019experimental}. 
Many COPs requiring self-interaction terms after mapping to the Ising model, including graph coloring, prime factorization, bin packing, etc., remain yet to be solved by dynamical Ising solvers.
Other unconventional approaches, including quantum 
and photonic implementations, generally utilize their unique physical behavior to directly represent the Ising models. 
However, many of them are challenging to integrate into silicon VLSI technologies. For example, the D-Wave quantum annealers proposed in \cite{albash2018demonstration, denchev2016computational, boixo2016computational} require expensive cryogenic cooling and exhibit limited connectivity between spins.
Optical Ising machine consumes extremely long optic fiber to implement the solver, making its integration highly challenging \cite{honjo2021100}.

In this article, we perform a hardware-algorithm co-design of a compute-in-memory (CiM) based annealer to efficiently solve QUBO formulations, thus the COPs, as shown in Fig.\ref{fig:motivation}\textbf{e}. 
The most well-known CiM hardware system is probably the crossbar array for acceleration of the vector-matrix multiplication (VMM), a core operation in neural networks \cite{sebastian2020memory}. 
In this scheme, the matrix is stored in memory, including volatile and nonvolatile memory (NVM), and the VMM computations are performed in CiM arrays without energy-consuming and slow data movement between memory and computing units, thus exhibiting superior energy efficiency.
Drawing inspiration from this, and recognizing that the QUBO formulation is composed of a vector-matrix-vector (VMV) multiplication as shown in Eq. \ref{equ:QUBO}, this article aims at expediting the in-situ VMV multiplication through CiM approach, thus accelerating solving COPs. Our CiM  annealer could potentially address the aforementioned challenges of digital annealers and dynamical Ising solvers. 
By storing the matrix in memory and performing the VMV multiplication directly in-memory, the CiM approach exhibits superior energy efficiency without suffering from the data movement bottleneck in the digital annealers.  
By performing the VMV multiplication in analog domain, CiM annealer 
is intrinsically robust against the noise and inaccuracy of the coupling matrix mapping. 
Since it does not rely on the overall system dynamics to solve the COPs, a large COP can be easily mapped across multiple chips, thus can be easily scalable. 
Lastly, the self-interaction term can also be easily implemented in the CiM array. Therefore, CiM approach when seamlessly integrated with efficient annealing algorithms could offer a powerful hardware platform for COPs.   

Here we propose to develop an ferroelectric field effect transistor (FeFET) based CiM crossbar array to accelerate VMV multiplications of QUBO, as shown in Fig.\ref{fig:motivation}\textbf{f}. FeFETs based on ferroelectric HfO\textsubscript{2} are a prime candidate technology platform to implement CiM system for in-situ VMV multiplication. First, it is naturally a three-terminal nonvolatile device, ideal for VMV multiplication, where the coupling matrix element can be stored in the polarization state of the FeFET and the two inputs (not necessarily identical) can be applied on the gate and drain, respectively. 
On the contrary, other two-terminal NVM based CiM system would require an VMM operation to calculate the intermediate result, and then apply another VMM to complete the VMV multiplication.
Second, HfO\textsubscript{2} based FeFET exhibits superior write energy efficiency with its electric field driven polarization switching mechanism, excellent scalability, and great CMOS compatibility \cite{schroeder2022fundamentals, salahuddin2018era}. Therefore, a compact single FeFET CiM array is developed in this work for the QUBO computations. 
The innovation of this work lies in: i) first proposal of a compact and effective 1FeFET CiM implementation for in-situ VMV multiplication  by exploiting the three-terminal structure and nonvolatile storage of FeFETs;
ii) proposing a lossless compression method for the QUBO formulation by capitalizing on the FeFET CiM array's capability to accommodate asymmetrical (non-identical) input vectors as shown in Fig.\ref{fig:motivation}\textbf{g}, thus significantly reducing the array size of the  crossbar and expanding the problem-solving capacity to larger scales;
iii) introducing a multi-epoch simulated annealing (MESA) algorithm to enhance the annealing process and improve the solution quality, which can quickly find the optimal solution of COPs via iterative QUBO computations; 
iv) first experimental demonstration of a FeFET CiM array to showcase its efficacy in accelerating QUBO computations and highly competitive performance against other hardware alternatives in solving complex COPs.

\section*{1FeFET1R based CiM architecture}
\label{sec:device}
Considering the great promise of FeFET crossbar array in accelerating VMV multiplication with both symmetric and asymmetric input vectors for COPs in QUBO formulation, FeFET CiM array is designed and experimentally demonstrated. 
Fig. \ref{fig:device} shows the cell and array design and experimental data illustrating the CiM hardware. The FeFET CiM chip is  integrated onto an industrial 28nm high-$\kappa$ metal gate FeFET technology platform \cite{trentzsch201628nm}. The device features an approximately 8nm doped HfO\textsubscript{2} as the ferroelectric layer, as shown in the schematic cross section and transmission electron microscopy (TEM) cross section in Fig.\ref{fig:device}\textbf{a}. The structural similarity of FeFET to standard logic transistor,  coupled with its CMOS compatibility and ultra-scalable nature of ferroelectric HfO\textsubscript{2}, enables the integration of FeFETs  with Si CMOS, which is leveraged in this work. 
For the demonstration, an 32$\times$32 FeFET array is designed, where the chip layout composed of array core, the word line (WL) driver, source line (SL)/data line (DL) driver, and the analog-to-digital converter (ADC) is shown in Fig.\ref{fig:device}\textbf{b}. 
The fabricated chip micrograph is shown in Fig.\ref{fig:device}\textbf{c}.

As shown in Fig.\ref{fig:device}\textbf{d}, our approach encodes the coupling matrix  element \textit{q} into the polarization states of the FeFET. By applying  inputs \textit{x} and \textit{y} to the FeFET's gate and drain, respectively, the resulting channel current $i_{DL}$ corresponds to the scalar product of these three, i.e., $i_{DL} = x\times q \times y$. 
Consequently, the core computation within VMV multiplication can be implemented with minimal overhead. This sets our approach aprat from other two-terminal NVM devices like memristors, which are limited to singular multiplications between the input and the stored values \cite{taoka2021simulated, misawa2022domain}. Fig.\ref{fig:device}\textbf{e} further shows the relationship between cell current and gate voltage (i.e., \textit{V}\textsubscript{WL}) for  two memory states across 60 distinct devices. 
The coupling matrix element is encoded as the polarization states, programmed via +4V/-4V, 1$\mu$s gate pulses, which induce the polarization to orient towards the channel/gate-metal, and hence set the threshold voltage (\textit{V}\textsubscript{TH}) of FeFET into the low-\textit{V}\textsubscript{TH} (i.e., \textit{q} = 1)/high-\textit{V}\textsubscript{TH} state (i.e., \textit{q} = 0), respectively. 
By choosing an appropriate read gate bias (i.e., input \textit{x}), the resultant cell current realizes the scalar product. 

While the design is compact and elegant, a potential challenge arises from the need to manage  FeFET variation, which can lead to compromised accuracy in  VMV multiplications. 
Despite ongoing improvements in materials and processes \cite{beyer2020fefet}, FeFET variation remains a significant factor in  CiM applications, 
as indicated in Fig.\ref{fig:device}\textbf{f}. 
In this work, we employ an 1FeFET1R cell structure as depicted in Fig.\ref{fig:device}\textbf{g} to effectively mitigate the device variations and enhance the accuracy of the VMV multiplication.
By incorporating a series resistor, 
the cell's ON current, regulated by the current limiter, becomes independent of the FeFET's ON current \cite{soliman2020ultra, saito2021analog}. 
Such structure ensures that
the presence of variation in \textit{V}\textsubscript{TH} does not manifest as variation in  the cell's ON current. 
As a proof of concept,  each FeFET is connected with a series resistor for the same group of 60 devices. Fig.\ref{fig:device}\textbf{h} shows the 1FeFET1R cell \textit{I}-\textit{V} characteristics, which exhibits the same  \textit{V}\textsubscript{TH} distribution as that in Fig.\ref{fig:device}\textbf{e}, while its ON current variation can be significantly suppressed as illustrated in Fig.\ref{fig:device}\textbf{i}. 
While there is a trade-off involving the reduction in the cell's ON current, the ON/OFF ratio  still exceeds 1000, ensuring that there should be no constraints on the practical array size.  
As a result, the 1FeFET1R CiM array is designed as illustrated in Fig. \ref{fig:device}\textbf{j}, where the resistor is implemented with a fully integrated MOSFET. 
The current within a column exhibits a linear relationship with the number of activated cells, corroborated across 20 different arrays as shown in Fig.\ref{fig:device}\textbf{k}.  
Therefore, it validates the linearity and functionality of the crossbar array, and also demonstrates the tightly controlled distribution of the output current.
These results lay a robust foundation for the acceleration of VMV multiplication in this work. 
A more detailed description of our chip measurement can be referred to Sec. \ref{sec:supp_chip}
of Supplementary Information.

\begin{figurehere}
	\centering
	\includegraphics [width=0.9\linewidth]{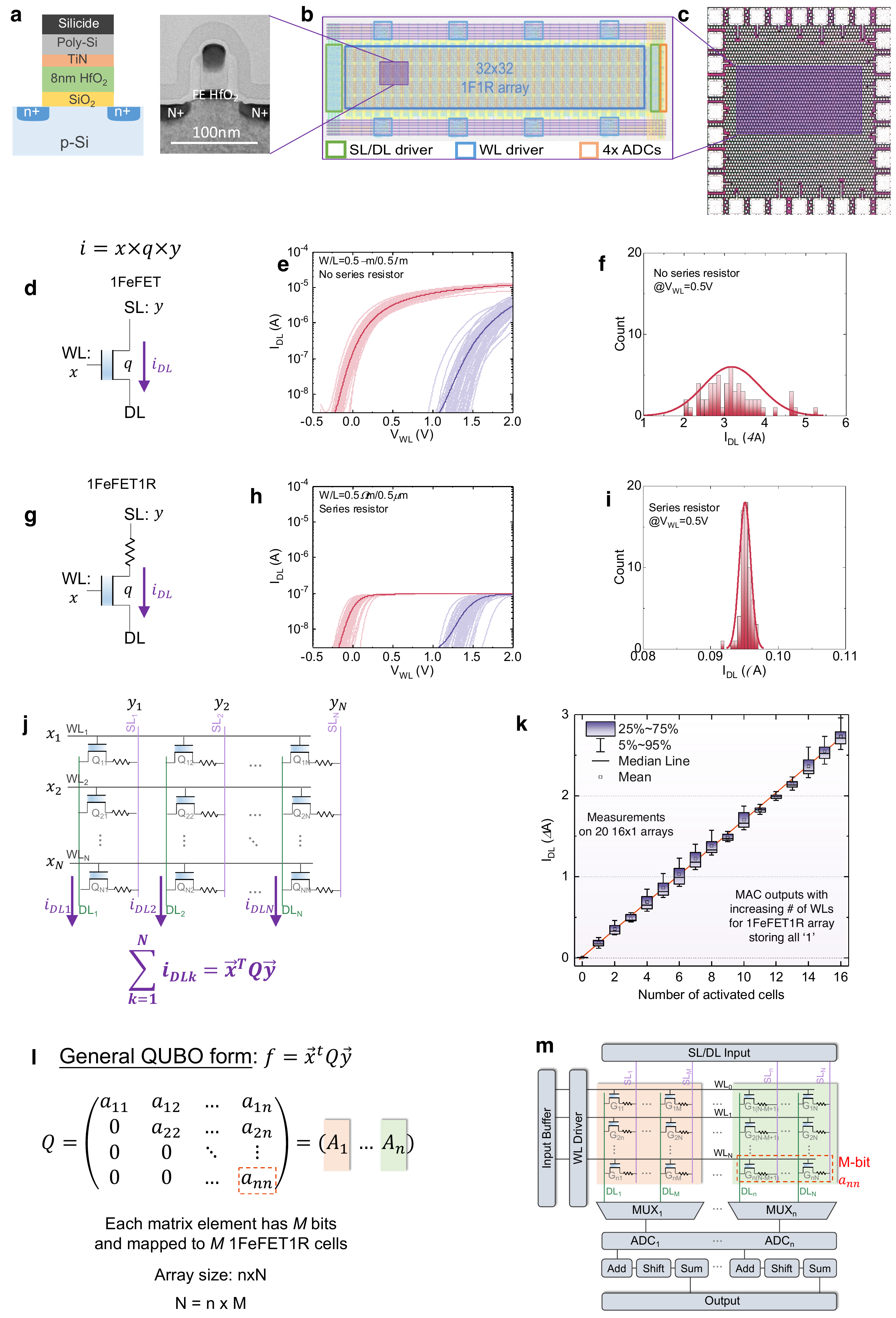}
 \caption{\textit{ \textbf{FeFET based CiM array for QUBO acceleration.}
 \textbf{a.} FeFET schematic and TEM cross section, featuring a 8nm doped HfO\textsubscript{2} s the ferroelectric. \textbf{b.} Layout of an 32$\times$32 FeFET array composed of core and peripherals. \textbf{c.} Micro-graph of the fabricated chip, where bond pads are visible.
 \textbf{d.} The current of a FeFET $i_{DL}$ corresponds to the scalar product of stored value $q$ (i.e., threshold voltage \textit{V}\textsubscript{TH}), and inputs $x$ and $y$, applied at the gate and drain, respectively. \textbf{e.} The \textit{I}\textsubscript{DL}-\textit{V}\textsubscript{WL} characteristics of 60 FeFETs for the two memory states. \textbf{f.} Significant ON current variation of FeFETs will result in compromised accuracy in VMV multiplications.
 \textbf{g.} An 1FeFET1R cell structure can suppress the \textit{I}\textsubscript{ON} variability. \textbf{h.} \textit{I}\textsubscript{DL}-\textit{V}\textsubscript{WL} of 60 1FeFET1R cells with 1M$\Omega$ resistor. \textbf{i.} A significantly narrower \textit{I}\textsubscript{ON} distribution for the 1FeFET1R cells, thus substantially enlarging the practical CiM array size.
 \textbf{j.} As a result, the 1FeFET1R CiM array implementing a general QUBO formulation is proposed. \textbf{k.} Measured column current shows a good linearity with respect to the number of activated cells in the column, thus promising for VMV multiplication.
 \textbf{l.} QUBO formulation mapping to \textbf{m.} FeFET based CiM array architecture.}}
	\label{fig:device}
\end{figurehere}

Lastly, we present the mapping of the generalized QUBO form, $\Vec{x}\ ^TQ\Vec{y}$, onto the 1FeFET1R CiM array, as illustrated in Fig.\ref{fig:device}\textbf{l} and \textbf{m}. The details of the general QUBO form are depicted in Fig.\ref{fig:device}\textbf{l}. 
The input vectors, $\Vec{x}$ and $\Vec{y}$, are mapped to the WL and SL inputs, respectively, as shown in Fig.\ref{fig:device}\textbf{m}.
This figure further shows the circuit implementation of the FeFET based crossbar array along with its associated peripheral circuits. 
The QUBO matrix is mapped onto the FeFET crossbar by storing  \textit{M}-bit precision matrix elements within \textit{M} 1FeFET1R cells. 
Each cell stores a single bit of the matrix element, therefore an \textit{n}$\times$\textit{n} QUBO matrix  corresponds to the implementation of \textit{n}$\times$\textit{N} cells, where \textit{N} is \textit{n}$\times$\textit{M}. 
To perform the VMV multiplication, the WL driver activates all rows of the FeFET crossbar, 
and the SLs of the columns are activated per 
the input $\Vec{y}$ (i.e., '1' indicates ON, and '0' indicates OFF).
The consecutive column outputs are directed to the column-shared analog-digital converters (ADCs), converted to digits, and further processed through Shift and Add units, generating the dot product between the stored multi-bit coupling vector and input vector $\Vec{x}$.
The final value of QUBO $\Vec{x}\ ^TQ\Vec{y}$ function  is then accumulated as the output of current iteration in the  annealing process, and stored in the output buffer.
In this way, our proposed CiM crossbar realizes the VMV multiplication directly by simply applying two input vectors within each iteration, and ultimately solves the QUBO formulation in Eq. \ref{equ:QUBO}, synergizing with the efficient annealing algorithms.

\section*{QUBO matrix lossless compression}
\label{sec:compression}

Mapping the QUBO formulation directly onto the FeFET crossbar CiM array with two identical or symmetrical input vectors applied on the WLs and SLs, respectively, has revealed a challenge in terms of low chip utilization. 
This issue stems from the inherent sparsity often observed in QUBO matrix converted from COPs.  Fig.\ref{fig:compress}\textbf{a} and \textbf{b} show the sparsity of QUBO matrix for graph coloring problems \cite{graphcolorcmu} and Max-Cut problems \cite{maxcutstanford}, respectively, across varying      problem instances with different node counts. 
Remarkably, the majority of the matrix elements (typically exceeding 85\%) assume zero values. 
Therefore, when directly mapping the QUBO matrix onto the FeFET crossbar array, a large portion of FeFET cells witin the  array are programmed to state '0' (i.e., high-\textit{V}\textsubscript{TH} state).
Although these 'OFF' cells do not actively participate in the  VMV multiplication during QUBO computation, they still incur additional hardware area overhead and contribute to leakage power consumption. 
With the expansion problem complexity and scale, the hardware size essential for accommodating the converted QUBO matrix exhibits quadratic growth  with the node count, thus leading to substantial hardware resources waste.
Such low hardware utilization therefore introduces formidable obstacles  to CiM annealers in solving larger-scale COPs efficiently.



To minimize the hardware inefficiencies stemming from  sparse matrix mapping, a lossless compression technique is proposed here, as illustrated in Fig.\ref{fig:compress}\textbf{c}. This approach entails pruning the sparse and symmetric QUBO matrix, 
originally of size $n\times n$ within the $\Vec{x}\ ^TQ\Vec{x}$ formulation, into  a more compact dense  matrix of size $p\times q$ within the $\Vec{x_h}\ ^TQ'\Vec{x_v}$ formulation, where $\Vec{x_h} \cup \Vec{x_v} = \Vec{x}$ and $p,q \leq n$.
This innovative technique achieves substantial chip size reduction by capitalizing on the distinctive attributes of 
the three-terminal FeFET crossbar CiM array, particularly when implementing the $\Vec{x}\ ^TQ\Vec{y}$ formulation, where $\Vec{x}$ and $\Vec{y}$ need not be symmetrical. 
Fig. \ref{fig:compress}\textbf{d} illustrates the methodology of QUBO matrix compression, elucidated through a concrete example, which consists of 4 steps:

\begin{figurehere}
	\centering
	\includegraphics [width=1.0\linewidth]{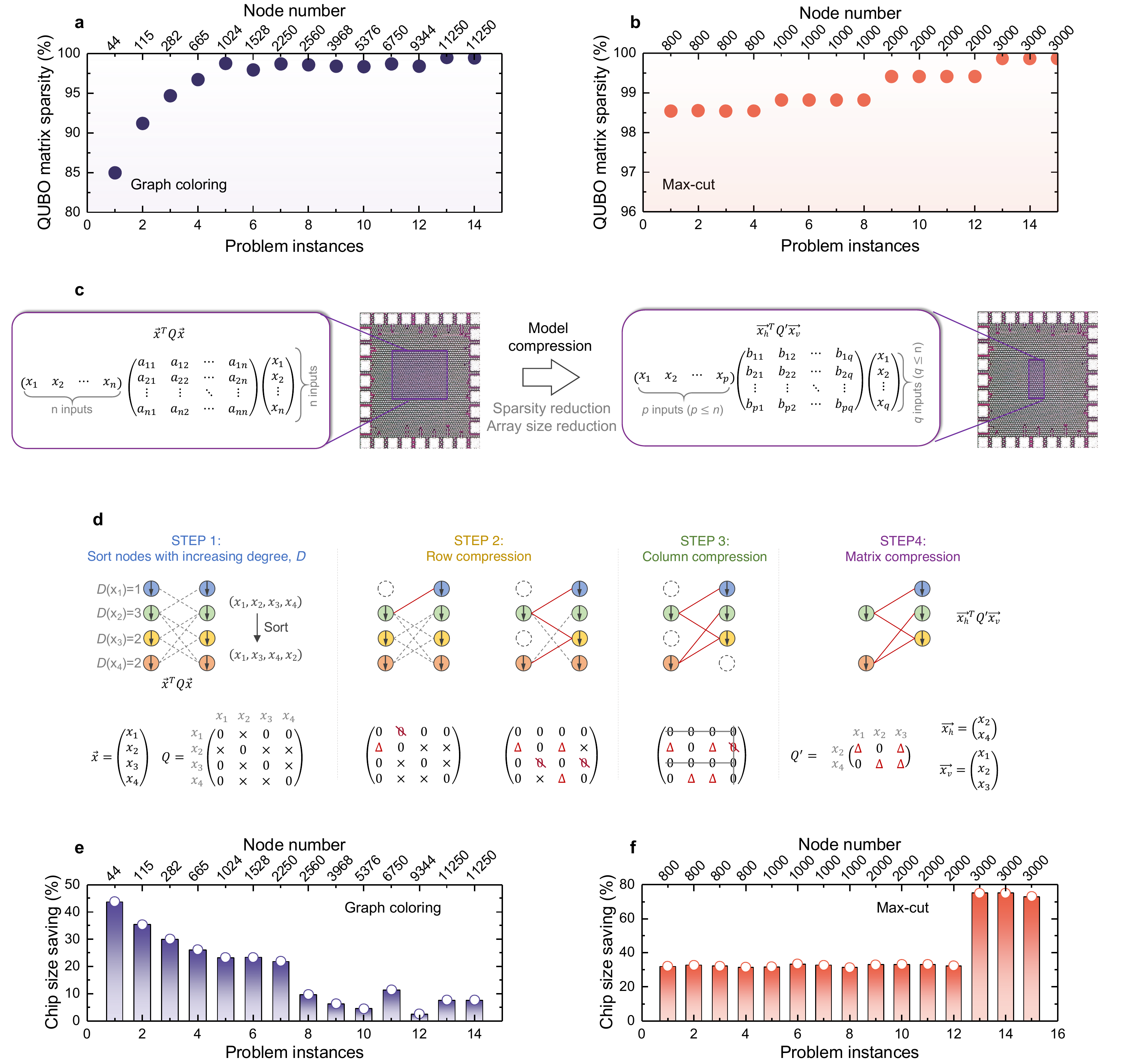}
	\caption{\textit{ \textbf{Proposed lossless compression of the QUBO matrix.} Conventional $\Vec{x}\ ^TQ\Vec{x}$ QUBO formulation for many COPs like \textbf{a.} graph coloring and \textbf{b.} Max-Cut is highly sparse. When directly mapped to a CiM array, significant portion of hardware could be wasted. \textbf{c.} A lossless compression method is proposed to convert a large scale $\Vec{x}\ ^TQ\Vec{x}$ formulation to a more compact and dense $\Vec{x_h}\ ^TQ'\Vec{x_v}$ formulation by leveraging the three-terminal FeFET based CiM array, where $\Vec{x} = \Vec{x_h} \cup \Vec{x_v}$. \textbf{d.} The conceptual flow of $\Vec{x}\ ^TQ\Vec{x}$ compression leveraging the symmetry of the \textit{Q} matrix. Chip size savings after compression for \textbf{e.} graph coloring and \textbf{f.} Max-Cut problems, presented in \textbf{a} and \textbf{b}, respectively.}}
	\label{fig:compress}
\end{figurehere}

\textbf{STEP 1}: The input vectors are organized in order of  the corresponding node degrees. Node degree represents the significance of the associated input variable in  QUBO formulation computation, gauging the extent of its involvement in nonzero scalar multiplications.
Pruning is initiated from  nodes with fewer connections.

\textbf{STEP 2}: Row compression of \textit{Q} matrix is carried out in the order of sorted input vector list obtained in STEP 1. For each selected input variable $x_i$  from the list, if the respective row in matrix \textit{Q} is compressible, every nonzero element within the row  is added to the element at its corresponding diagonal position, capitalizing on the symmetrical nature of $\Vec{x}\ ^TQ\Vec{x}$ formulation. 
Subsequently,
the elements in the compressed row are  set to zero, with rows containing the updated diagonal elements  marked as incompressible to ensure lossless compression.

\textbf{STEP 3}: Column compression mirrors the operation conducted in STEP 2, with nonzero elements within the compressed column added to the their respective diagonal elements,  then set to zero.
Columns containing updated diagonal elements are labeled as incompressible.

\textbf{STEP 4}: The QUBO matrix's compressed rows and columns, along with their corresponding variables in the  input vectors, are eliminated, yielding the compressed QUBO matrix along with the compressed input vectors.

A detailed description of the compression methodology is featured in Sec.\ref{sec:supp_compression} of Supplementary Information.
As a result, redundant rows and columns of QUBO matrix are removed, reducing  crossbar array size required to implement the $\Vec{x}\ ^TQ\Vec{x}$  QUBO formulation  without sacrificing accuracy.
The compressed QUBO matrix is  mapped onto the FeFET crossbar for the iterative annealing of QUBO formalized COPs.
The binary variable vectors $\Vec{x_h}$ and $\Vec{x_v}$ associated with the compressed QUBO formulation are applied to the WLs and SLs of the FeFET CiM array, respectively.
In this way, the proposed compression approach enhances the scalability of  CiM hardware, thereby scaling up the capacity for solving larger-scale COPs.

The efficacy of the proposed compression technique has been evaluated. For the same problem instances of graph coloring and Max-Cut COPs, as analyzed in Fig.\ref{fig:compress}\textbf{a} and \textbf{b}, respectively, the corresponding chip size reduction percentages are elucidated in Fig.\ref{fig:compress}\textbf{e} and \textbf{f}. 
These results demonstrate that the compression method yields substantial savings in chip size. 
Note that the extent of chip size reduction does not necessarily correlates with the node count in a problem. This is because that
the distribution of nonzero elements within the QUBO matrix significantly affects the impact of the compression method. 
For instance, if all nonzero 
elements aggregate within a single row, the QUBO matrix could be compressed to just one row, yielding high chip size savings. Conversely, if each row contains only one nonzero element, and these nonzero elements are at different columns, compression of the QUBO matrix might not be feasible, even if it is sparse.

\section*{Solving COPs with multi-epoch simulated annealing}
\label{sec:evaluation}


Previously developed FeFET CiM array demonstrates its capability to accelerate the computation of the QUBO formulation, which matches with annealing algorithms for solving COPs.
That is, the configurations or solutions corresponding to the minimal QUBO energy are sought via an iterative annealing procedure. 
Simulated annealing (SA) algorithms were introduced to address the problem of local minimum trapping  during the annealing process.
The energy of the objective function, as computed by the configurations in current iteration, is  compared with the energy state corresponding to current solution. 
If the computed energy is lower, 
the solution configurations are updated with the corresponding variable configurations. Conversely, if the  energy is higher,  the  update retains a probability proportional to the temperature.
Nonetheless, conventional SA has demonstrated suboptimal performance in handling large-scale COPs \cite{hong2021memory}.
To accelerate the SA process while still ensuring high probability of finding optimal solutions, a multi-epoch simulated annealing (MESA) algorithm is herein proposed.  

Fig.\ref{fig:result}\textbf{a} illustrates the MESA process.
For each epoch,  an optimal solution ($\Vec{x}_{hopt}, \Vec{x}_{vopt}$) and its associated QUBO energy $E_{opt}$ are defined and sustained throughout the epoch. 
This records the lowest energy state attainable by the system, given the input configuration and the energy initialized to the optimal solution from the previous epoch. 
The QUBO energy, $E_{new}$, is calculated using the CiM hardware, as  previously detailed. 
If $E_{new}$ is lower than the energy $E_{o}$ of the last iteration, indicating a progression toward a lower energy landscape, this QUBO energy and its associated  variable solution ($\Vec{x_h}, \Vec{x_v}$) are accepted.
The optimal solution ($\Vec{x}_{hopt}, \Vec{x}_{vopt}$) along with its energy value $E_{opt}$ within this epoch are either updated or maintained, depending on the comparison between $E_{new}$ and the optimal solution. 
Should $E_{new}$ closely approximate $E_{o}$, indicating that the system is trapped at a local minimum,
the energy and its corresponding variable solution remain unaltered, and the trap count is updated.
If $E_{new}$ is notably larger than the energy $E_o$, the system has a probability closely related to the temperature $T$ to accept the variable solution, allowing a chance to escape from local minimum.
Subsequently, the system introduces random perturbations 
by flipping a few bits in the input vectors, and proceeds to the next iteration.
When the system is trapped at a local minimum, where the energy trajectory remains stagnant  for  a predefined period $Count_{max}$, the epoch concludes,  the temperature resets, and  a fresh epoch commences. 
As a result, the length of each epoch  is dynamically adjusted according to the  system's advancement.
The input configuration and its corresponding energy at the beginning of a new SA epoch are initialized based on the optimal solution and the energy recorded during the last epoch. 
Such initialization ensures  the continuous convergence  of MESA toward  lower energy states.

\begin{figurehere}
	\centering
	\includegraphics [width=0.95\linewidth]{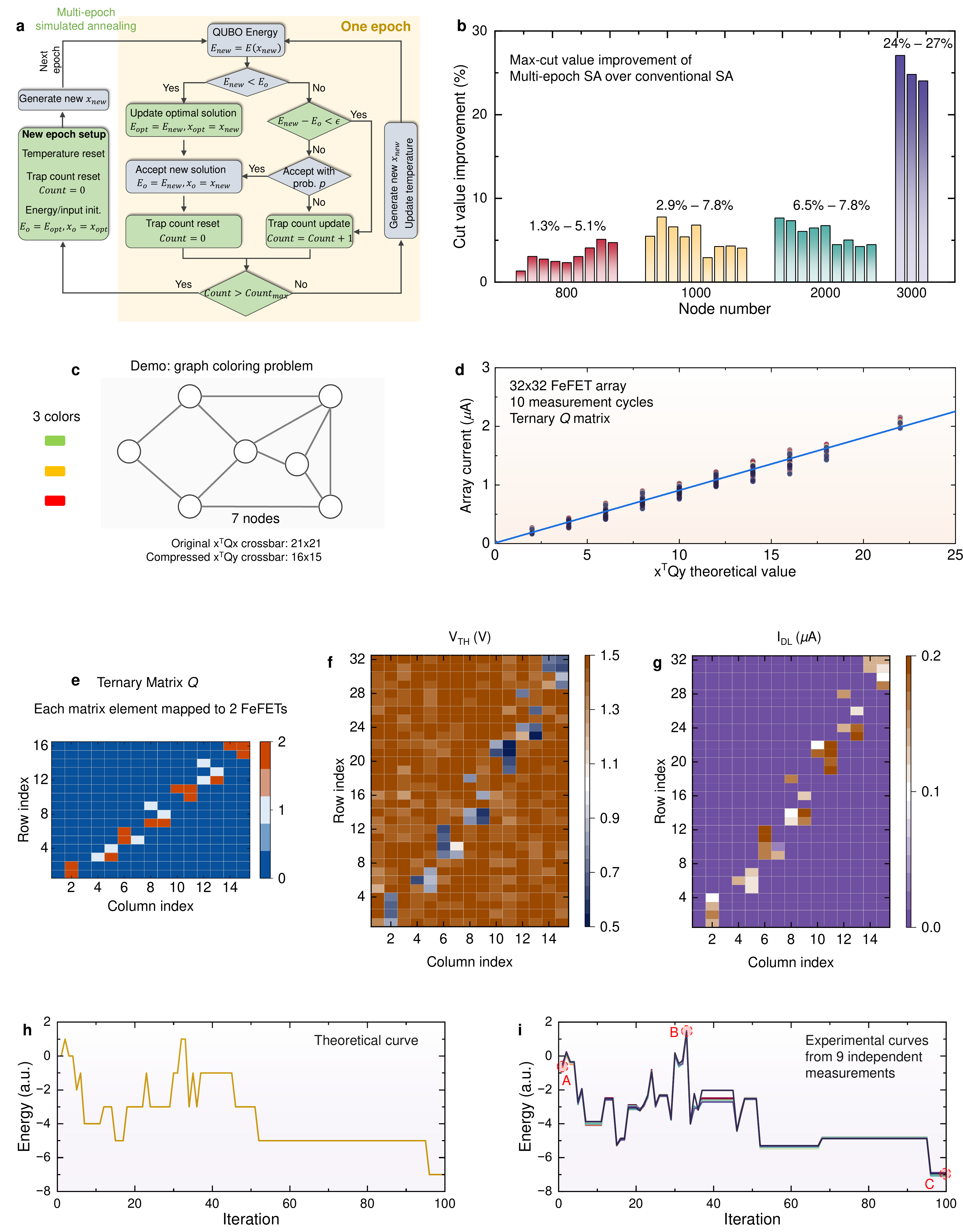}
	\caption{\textit{ \textbf{Demonstration of a new simulated annealing algorithm and its hardware acceleration using FeFET CiM array.}
    \textbf{a.} Proposed multi-epoch simulated annealing (MESA) algorithm for solving complex COPs.
    \textbf{b.} Results of 30 Max-Cut problems demonstrate that MESA outperforms conventional SA in the quality of solution.
    \textbf{c.} A toy example of graph coloring for hardware implementation, which consists of 7 nodes and 3 colors to paint. After compression, it requires a ternary CiM array of 16x15 to map.
    \textbf{d.} When mapped to a FeFET CiM array, the measured array current is linearly proportional to the theoretical $\Vec{x}\ ^TQ\Vec{y}$ value, demonstrating the feasibility of performing MESA in hardware. 
    \textbf{e.} The ternary matrix \textit{Q} corresponding to the toy graph coloring problem in \textbf{c}.
    \textbf{f.} Using 2 FeFETs to represent each ternary \textit{Q} matrix element, the programmed FeFET \textit{V}\textsubscript{TH} map of the corresponding 32x15 array. \textbf{g.} The corresponding cell current. \textbf{h.} The theoretical energy as a function of annealing iteration. \textbf{i.} Experimentally measured energy as a function of annealing iteration, showing successful operation of the hardware in implementing the annealing algorithm.
	}}
	\label{fig:result}
\end{figurehere}



Fig. \ref{fig:result}\textbf{b} shows the COP solving capability of MESA over that of conventional SA in identifying optimal solutions for prevalent Max-Cut problems.
It can be seen  that MESA outperforms the conventional SA across all Max-Cut problems,  ranging in size from 800  to 3000 nodes, encompassing  a search space spanning $2^{800}$ to $2^{3000}$.
As the node count increases, i.e., the problem complexity grows, MESA yields superior solutions to SA, boasting an improvement of up to 27\% in cut value when addressing 3000-node problems. 
With the dataset tested, Fig.\ref{fig: supp MESA} shows that MESA consistently outperforms conventional SA in terms of success rate and time-to-solution.
Yet, the capability of all the CiM SA solver relies not only on the algorithm, but also the precision of the hardware especially when mapping the QUBO matrix.
Fig.\ref{fig:SA 323} demonstrates the impact of the QUBO matrix precision in solving the prime factorization problem (PFP), as an example.  
A single MESA epoch of solving the PFP is studied. 
As expected, the higher the precision of QUBO matrix, the higher the success rates of MESA in finding the solution. This is because higher mapping precision yields a more accurate representation of the energy landscape, as shown in Fig.\ref{fig: 323 precision}, which however incurs more hardware costs in terms of analog-digital-converter and array size.

To demonstrate the capability of developed FeFET CiM array in accelerating COPs, a toy example of graph coloring, as shown in Fig.\ref{fig:result}\textbf{c}, is evaluated. To fit the problem into the developed 32$\times$32 FeFET array, the example consists 7 nodes and 3 colors to be assigned. The initial QUBO formulation prior to compression necessitates an 21$\times$21 array (i.e., each node can be any of the three color, thus total 21 input variables, per the QUBO conversion described in Sec.\ref{sec:conversion} of Supplementary Information), whereas the compressed formulation can be implemented on an 16$\times$15 array, resulting in a notable 1.84$\times$ reduction in chip size. 
In this example, the matrix \textit{Q} is ternary (i.e., has value of 0, 1, 2). For experimental demonstration, 2 FeFETs are used to represent each matrix element. Three FeFET CiM array dies (see Fig. \ref{fig:device}\textbf{c} for the chip photo) have been employed to evaluate the QUBO formulation. The capability of the chip in realizing the intended $\Vec{x}\ ^TQ\Vec{y}$ computation is demonstrated in Fig.\ref{fig:result}\textbf{d}, where the measured array total current shows a linear dependence on the theoretical $\Vec{x}\ ^TQ\Vec{y}$ value. Building upon this capability,  MESA is  performed on the chip. Fig.\ref{fig:result}\textbf{e} shows the ternary \textit{Q} matrix corresponding to the graph coloring problem in Fig.\ref{fig:result}\textbf{c}. The corresponding 32$\times$15 FeFET array is then programmed. Fig.\ref{fig:result}\textbf{f} and \textbf{g} show the \textit{V}\textsubscript{TH} and measured cell current, respectively, demonstrating successful mapping of the matrix \textit{Q}. 

Fig.\ref{fig:result}\textbf{h} and \textbf{i} show the theoretical energy and experimentally measured energy evolution with annealing iterations. 
The experimental measurements are conducted on 3 separate dies. 
Fig. \ref{fig:result}\textbf{i} shows 9 separate measurements on one of the die, where for each measurement, the FeFET CiM array is erased and programmed again with the same QUBO matrix, and  MESA is executed. 
All of the measured curves consistently align with the theoretical curve, validating the capability and robustness of the proposed FeFET CiM systems in performing MESA to solve COPs.
More experimental measurement results can be found in Fig.\ref{fig: die all}.
Fig.\ref{fig: graphcolor abc}
showcases the graph coloring configurations during annealing, highlighted at different iteration steps shown in Fig.\ref{fig:result}\textbf{i}, i.e., the beginning (A), midpoint (B), and end (C) of the evolution process.
Initially, errors such as multiple colors attributed to a single node, identical colors assigned to adjacent nodes, or uncolored nodes could occur, which correspond to high QUBO energy states like point A and B. In these cases, the algorithm is still exploring the solution space with the constraints loosely enforced.
As annealing proceeds, the proposed approach can ultimately find the optimal solution with all the constraints satisfied.
\section*{Conclusion}
\label{sec:conclusion}

We proposed a comprehensive hardware-algorithm co-design framework for solving the complex COPs efficiently. 
The proposed approach comprises a FeFET based CiM array that accelerates the critical in-situ VMV multiplications within the QUBO formulation. Additionally,  the proposed QUBO matrix compression technique significantly improves the 
chip utilization, thereby
enhancing the problem solving capability of the hardware when addressing larger COPs.
Complementing this, our multi-epoch based SA algorithm optimizes the proposed solver's ability to converge and reach optimal solutions within a shorter time period.
Both the simulation and experimental measurements on fabricated prototypes validate the problem-solving capability of the proposed approach.
The solver summary in Table. \ref{table:evaluation} demonstrates that  
the proposed framework 
can outperform solvers for COPs commonly showcased in prior works by leveraging the VMV acceleration and QUBO compression techniques. 
Remarkably, the proposed framework showcases robust COP-solving capability and exhibits wide applicability to a broad spectrum of COPS that can be transformed to QUBO formulation.

\begin{table*}

\caption{Summary of QUBO Solvers}
\label{table:evaluation}
\centering
\resizebox{\columnwidth}{!}{
\begin{tabular}{| c | c | c | c | c | c| c|}
\hline\hline
Reference & \cite{cai2020power} & \cite{yang2020transiently}& \cite{shin2018hardware}& \cite{mahmoodi2019analog}& \cite{hong2021memory} & This work\\
\hline
\multirow{3}{*}{Problem} & \multirow{3}{*}{Max-Cut} &\multirow{3}{*}{Max-Cut} &\multirow{2}{*}{Spin} &\multirow{2}{*}{Graph} &\multirow{2}{*}{Traveling}&Graph Coloring/\\
 & & &\multirow{2}{*}{Glass} &\multirow{2}{*}{Partion} &\multirow{2}{*}{Salesman } &Max-Cut/ \\
 & & & & & &Prime Factorization\\
\hline
QUBO Matrix  & \multirow{2}{*}{No} & \multirow{2}{*}{No} & \multirow{2}{*}{No} & \multirow{2}{*}{No} & \multirow{2}{*}{No} & \multirow{2}{*}{Yes}\\
Compression&  &  &  &  &  & \\
\hline
Method
 &  SA & Chaotic SA & SA &  SA &  Multi-step SA & MESA\\
\hline
Hardware  & memristor & memristor & RRAM & RRAM & RRAM & FeFET\\
Implementation&based crossbar  &based crossbar  &based crossbar  &based crossbar  &based crossbar  & based crossbar$^\dagger$\\

\hline

Hardware Acceleration$^\star$ & VM & VM & VM & VM & VM & VMV\\
\hline

Hardware Size & 60x60 & 2x2 & 11x3 & 64x64 & 1024x1152 & 32x32\\
 \hline
 Problem Size& 60 node  & 5 node  & 15 node & 6 node  & 100 node  & 21 node\\
 \hline

\end{tabular}
}

 \begin{flushleft}
 \scriptsize
$\star$: VM denotes vector-matrix multiplication, VMV denots vector-matrix-vector multiplication.\\
$^\dagger$: The first one known to us.

 \end{flushleft}
\end{table*}

\section*{Data availability}
\label{sec:data}

The data that support the plots within this paper and other findings of this study are
available from the corresponding author on reasonable request.

\section*{Code availability}
\label{sec:data}

Custom code used in this study is available from the corresponding authors upon reasonable request.

\bibliography{ref}

\begin{thebibliography}{10}
\expandafter\ifx\csname url\endcsname\relax
  \def\url#1{\texttt{#1}}\fi
\expandafter\ifx\csname urlprefix\endcsname\relax\def\urlprefix{URL }\fi
\providecommand{\bibinfo}[2]{#2}
\providecommand{\eprint}[2][]{\url{#2}}

\bibitem{yu2013industrial}
\bibinfo{author}{Yu, G.}
\newblock \emph{\bibinfo{title}{Industrial applications of combinatorial
  optimization}}, vol.~\bibinfo{volume}{16} (\bibinfo{publisher}{Springer
  Science \& Business Media}, \bibinfo{year}{2013}).

\bibitem{paschos2014applications}
\bibinfo{author}{Paschos, V.~T.}
\newblock \emph{\bibinfo{title}{Applications of combinatorial optimization}},
  vol.~\bibinfo{volume}{3} (\bibinfo{publisher}{John Wiley \& Sons},
  \bibinfo{year}{2014}).

\bibitem{naseri2020application}
\bibinfo{author}{Naseri, G.} \& \bibinfo{author}{Koffas, M.~A.}
\newblock \bibinfo{title}{Application of combinatorial optimization strategies
  in synthetic biology}.
\newblock \emph{\bibinfo{journal}{Nature communications}}
  \textbf{\bibinfo{volume}{11}}, \bibinfo{pages}{2446} (\bibinfo{year}{2020}).

\bibitem{barahona1988application}
\bibinfo{author}{Barahona, F.}, \bibinfo{author}{Gr{\"o}tschel, M.},
  \bibinfo{author}{J{\"u}nger, M.} \& \bibinfo{author}{Reinelt, G.}
\newblock \bibinfo{title}{An application of combinatorial optimization to
  statistical physics and circuit layout design}.
\newblock \emph{\bibinfo{journal}{Operations Research}}
  \textbf{\bibinfo{volume}{36}}, \bibinfo{pages}{493--513}
  (\bibinfo{year}{1988}).

\bibitem{markov2014limits}
\bibinfo{author}{Markov, I.~L.}
\newblock \bibinfo{title}{Limits on fundamental limits to computation}.
\newblock \emph{\bibinfo{journal}{Nature}} \textbf{\bibinfo{volume}{512}},
  \bibinfo{pages}{147--154} (\bibinfo{year}{2014}).

\bibitem{markov2013know}
\bibinfo{author}{Markov, I.~L.}
\newblock \bibinfo{title}{Know your limits (review of" limits to parallel
  computation: p-completeness theory"; greenlaw, r., et al; 1995)[book
  review]}.
\newblock \emph{\bibinfo{journal}{IEEE Design \& Test}}
  \textbf{\bibinfo{volume}{30}}, \bibinfo{pages}{78--83}
  (\bibinfo{year}{2013}).

\bibitem{greenlaw1995limits}
\bibinfo{author}{Greenlaw, R.}, \bibinfo{author}{Hoover, H.~J.} \&
  \bibinfo{author}{Ruzzo, W.~L.}
\newblock \emph{\bibinfo{title}{Limits to parallel computation: P-completeness
  theory}} (\bibinfo{publisher}{Oxford University Press on Demand},
  \bibinfo{year}{1995}).

\bibitem{lucas2014ising}
\bibinfo{author}{Lucas, A.}
\newblock \bibinfo{title}{Ising formulations of many np problems}.
\newblock \emph{\bibinfo{journal}{Frontiers in physics}}
  \textbf{\bibinfo{volume}{2}}, \bibinfo{pages}{5} (\bibinfo{year}{2014}).

\bibitem{mohseni2022ising}
\bibinfo{author}{Mohseni, N.}, \bibinfo{author}{McMahon, P.~L.} \&
  \bibinfo{author}{Byrnes, T.}
\newblock \bibinfo{title}{Ising machines as hardware solvers of combinatorial
  optimization problems}.
\newblock \emph{\bibinfo{journal}{Nature Reviews Physics}}
  \textbf{\bibinfo{volume}{4}}, \bibinfo{pages}{363--379}
  (\bibinfo{year}{2022}).

\bibitem{date2021qubo}
\bibinfo{author}{Date, P.}, \bibinfo{author}{Arthur, D.} \&
  \bibinfo{author}{Pusey-Nazzaro, L.}
\newblock \bibinfo{title}{Qubo formulations for training machine learning
  models}.
\newblock \emph{\bibinfo{journal}{Scientific reports}}
  \textbf{\bibinfo{volume}{11}}, \bibinfo{pages}{10029} (\bibinfo{year}{2021}).

\bibitem{zaman2021pyqubo}
\bibinfo{author}{Zaman, M.}, \bibinfo{author}{Tanahashi, K.} \&
  \bibinfo{author}{Tanaka, S.}
\newblock \bibinfo{title}{Pyqubo: Python library for mapping combinatorial
  optimization problems to qubo form}.
\newblock \emph{\bibinfo{journal}{IEEE Transactions on Computers}}
  \textbf{\bibinfo{volume}{71}}, \bibinfo{pages}{838--850}
  (\bibinfo{year}{2021}).

\bibitem{glover2018tutorial}
\bibinfo{author}{Glover, F.}, \bibinfo{author}{Kochenberger, G.} \&
  \bibinfo{author}{Du, Y.}
\newblock \bibinfo{title}{A tutorial on formulating and using qubo models}.
\newblock \emph{\bibinfo{journal}{arXiv preprint arXiv:1811.11538}}
  (\bibinfo{year}{2018}).

\bibitem{glover2022quantum}
\bibinfo{author}{Glover, F.}, \bibinfo{author}{Kochenberger, G.},
  \bibinfo{author}{Hennig, R.} \& \bibinfo{author}{Du, Y.}
\newblock \bibinfo{title}{Quantum bridge analytics i: a tutorial on formulating
  and using qubo models}.
\newblock \emph{\bibinfo{journal}{Annals of Operations Research}}
  \textbf{\bibinfo{volume}{314}}, \bibinfo{pages}{141--183}
  (\bibinfo{year}{2022}).

\bibitem{yamamoto20207}
\bibinfo{author}{Yamamoto, K.}, \bibinfo{author}{Ando, K.},
  \bibinfo{author}{Mertig, N.}, \bibinfo{author}{Takemoto, T.},
  \bibinfo{author}{Yamaoka, M.}, \bibinfo{author}{Teramoto, H.},
  \bibinfo{author}{Sakai, A.}, \bibinfo{author}{Takamaeda-Yamazaki, S.} \&
  \bibinfo{author}{Motomura, M.}
\newblock \bibinfo{title}{Statica: A 512-spin 0.25 m-weight full-digital
  annealing processor with a near-memory all-spin-updates-at-once architecture
  for combinatorial optimization with complete spin-spin interactions}.
\newblock In \emph{\bibinfo{booktitle}{2020 IEEE International Solid-State
  Circuits Conference-(ISSCC)}}, \bibinfo{pages}{138--140}
  (\bibinfo{organization}{IEEE}, \bibinfo{year}{2020}).

\bibitem{katsuki2022fast}
\bibinfo{author}{Katsuki, K.}, \bibinfo{author}{Shin, D.},
  \bibinfo{author}{Onizawa, N.} \& \bibinfo{author}{Hanyu, T.}
\newblock \bibinfo{title}{Fast solving complete 2000-node optimization using
  stochastic-computing simulated annealing}.
\newblock In \emph{\bibinfo{booktitle}{2022 29th IEEE International Conference
  on Electronics, Circuits and Systems (ICECS)}}, \bibinfo{pages}{1--4}
  (\bibinfo{organization}{IEEE}, \bibinfo{year}{2022}).

\bibitem{onizawa2023local}
\bibinfo{author}{Onizawa, N.}, \bibinfo{author}{Kuroki, K.},
  \bibinfo{author}{Shin, D.} \& \bibinfo{author}{Hanyu, T.}
\newblock \bibinfo{title}{Local energy distribution based hyperparameter
  determination for stochastic simulated annealing}.
\newblock \emph{\bibinfo{journal}{arXiv preprint arXiv:2304.11839}}
  (\bibinfo{year}{2023}).

\bibitem{takemoto20214}
\bibinfo{author}{Takemoto, T.}, \bibinfo{author}{Yamamoto, K.},
  \bibinfo{author}{Yoshimura, C.}, \bibinfo{author}{Hayashi, M.},
  \bibinfo{author}{Tada, M.}, \bibinfo{author}{Saito, H.},
  \bibinfo{author}{Mashimo, M.} \& \bibinfo{author}{Yamaoka, M.}
\newblock \bibinfo{title}{A 144kb annealing system composed of 9$\times$ 16kb
  annealing processor chips with scalable chip-to-chip connections for
  large-scale combinatorial optimization problems}.
\newblock In \emph{\bibinfo{booktitle}{2021 IEEE International Solid-State
  Circuits Conference (ISSCC)}}, vol.~\bibinfo{volume}{64},
  \bibinfo{pages}{64--66} (\bibinfo{organization}{IEEE}, \bibinfo{year}{2021}).

\bibitem{moy20221}
\bibinfo{author}{Moy, W.}, \bibinfo{author}{Ahmed, I.}, \bibinfo{author}{Chiu,
  P.-w.}, \bibinfo{author}{Moy, J.}, \bibinfo{author}{Sapatnekar, S.~S.} \&
  \bibinfo{author}{Kim, C.~H.}
\newblock \bibinfo{title}{A 1,968-node coupled ring oscillator circuit for
  combinatorial optimization problem solving}.
\newblock \emph{\bibinfo{journal}{Nature Electronics}}
  \textbf{\bibinfo{volume}{5}}, \bibinfo{pages}{310--317}
  (\bibinfo{year}{2022}).

\bibitem{ahmed2021probabilistic}
\bibinfo{author}{Ahmed, I.}, \bibinfo{author}{Chiu, P.-W.},
  \bibinfo{author}{Moy, W.} \& \bibinfo{author}{Kim, C.~H.}
\newblock \bibinfo{title}{A probabilistic compute fabric based on coupled ring
  oscillators for solving combinatorial optimization problems}.
\newblock \emph{\bibinfo{journal}{IEEE Journal of Solid-State Circuits}}
  \textbf{\bibinfo{volume}{56}}, \bibinfo{pages}{2870--2880}
  (\bibinfo{year}{2021}).

\bibitem{dutta2021ising}
\bibinfo{author}{Dutta, S.}, \bibinfo{author}{Khanna, A.},
  \bibinfo{author}{Assoa, A.}, \bibinfo{author}{Paik, H.},
  \bibinfo{author}{Schlom, D.~G.}, \bibinfo{author}{Toroczkai, Z.},
  \bibinfo{author}{Raychowdhury, A.} \& \bibinfo{author}{Datta, S.}
\newblock \bibinfo{title}{An ising hamiltonian solver based on coupled
  stochastic phase-transition nano-oscillators}.
\newblock \emph{\bibinfo{journal}{Nature Electronics}}
  \textbf{\bibinfo{volume}{4}}, \bibinfo{pages}{502--512}
  (\bibinfo{year}{2021}).

\bibitem{roychowdhury2021bistable}
\bibinfo{author}{Roychowdhury, J.}
\newblock \bibinfo{title}{Bistable latch ising machines}.
\newblock In \emph{\bibinfo{booktitle}{Unconventional Computation and Natural
  Computation: 19th International Conference, UCNC 2021, Espoo, Finland,
  October 18--22, 2021, Proceedings 19}}, \bibinfo{pages}{131--148}
  (\bibinfo{organization}{Springer}, \bibinfo{year}{2021}).

\bibitem{mallick2023cmos}
\bibinfo{author}{Mallick, A.}, \bibinfo{author}{Zhao, Z.},
  \bibinfo{author}{Bashar, M.~K.}, \bibinfo{author}{Alam, S.},
  \bibinfo{author}{Islam, M.~M.}, \bibinfo{author}{Xiao, Y.},
  \bibinfo{author}{Xu, Y.}, \bibinfo{author}{Aziz, A.},
  \bibinfo{author}{Narayanan, V.}, \bibinfo{author}{Ni, K.} \emph{et~al.}
\newblock \bibinfo{title}{Cmos-compatible ising machines built using bistable
  latches coupled through ferroelectric transistor arrays}.
\newblock \emph{\bibinfo{journal}{Scientific reports}}
  \textbf{\bibinfo{volume}{13}}, \bibinfo{pages}{1515} (\bibinfo{year}{2023}).

\bibitem{afoakwa2021brim}
\bibinfo{author}{Afoakwa, R.}, \bibinfo{author}{Zhang, Y.},
  \bibinfo{author}{Vengalam, U. K.~R.}, \bibinfo{author}{Ignjatovic, Z.} \&
  \bibinfo{author}{Huang, M.}
\newblock \bibinfo{title}{Brim: bistable resistively-coupled ising machine}.
\newblock In \emph{\bibinfo{booktitle}{2021 IEEE International Symposium on
  High-Performance Computer Architecture (HPCA)}}, \bibinfo{pages}{749--760}
  (\bibinfo{organization}{IEEE}, \bibinfo{year}{2021}).

\bibitem{pierangeli2019large}
\bibinfo{author}{Pierangeli, D.}, \bibinfo{author}{Marcucci, G.} \&
  \bibinfo{author}{Conti, C.}
\newblock \bibinfo{title}{Large-scale photonic ising machine by spatial light
  modulation}.
\newblock \emph{\bibinfo{journal}{Physical review letters}}
  \textbf{\bibinfo{volume}{122}}, \bibinfo{pages}{213902}
  (\bibinfo{year}{2019}).

\bibitem{honjo2021100}
\bibinfo{author}{Honjo, T.}, \bibinfo{author}{Sonobe, T.},
  \bibinfo{author}{Inaba, K.}, \bibinfo{author}{Inagaki, T.},
  \bibinfo{author}{Ikuta, T.}, \bibinfo{author}{Yamada, Y.},
  \bibinfo{author}{Kazama, T.}, \bibinfo{author}{Enbutsu, K.},
  \bibinfo{author}{Umeki, T.}, \bibinfo{author}{Kasahara, R.} \emph{et~al.}
\newblock \bibinfo{title}{100,000-spin coherent ising machine}.
\newblock \emph{\bibinfo{journal}{Science advances}}
  \textbf{\bibinfo{volume}{7}}, \bibinfo{pages}{eabh0952}
  (\bibinfo{year}{2021}).

\bibitem{yamamoto2020coherent}
\bibinfo{author}{Yamamoto, Y.}, \bibinfo{author}{Leleu, T.},
  \bibinfo{author}{Ganguli, S.} \& \bibinfo{author}{Mabuchi, H.}
\newblock \bibinfo{title}{Coherent ising machines—quantum optics and neural
  network perspectives}.
\newblock \emph{\bibinfo{journal}{Applied Physics Letters}}
  \textbf{\bibinfo{volume}{117}}, \bibinfo{pages}{160501}
  (\bibinfo{year}{2020}).

\bibitem{mcmahon2016fully}
\bibinfo{author}{McMahon, P.~L.}, \bibinfo{author}{Marandi, A.},
  \bibinfo{author}{Haribara, Y.}, \bibinfo{author}{Hamerly, R.},
  \bibinfo{author}{Langrock, C.}, \bibinfo{author}{Tamate, S.},
  \bibinfo{author}{Inagaki, T.}, \bibinfo{author}{Takesue, H.},
  \bibinfo{author}{Utsunomiya, S.}, \bibinfo{author}{Aihara, K.},
  \bibinfo{author}{Byer, R.~L.}, \bibinfo{author}{Fejer, M.~M.},
  \bibinfo{author}{Mabuchi, H.} \& \bibinfo{author}{Yamamoto, Y.}
\newblock \bibinfo{title}{A fully programmable 100-spin coherent ising machine
  with all-to-all connections}.
\newblock \emph{\bibinfo{journal}{Science}} \textbf{\bibinfo{volume}{354}},
  \bibinfo{pages}{614--617} (\bibinfo{year}{2016}).

\bibitem{bohm2019poor}
\bibinfo{author}{B{\"o}hm, F.}, \bibinfo{author}{Verschaffelt, G.} \&
  \bibinfo{author}{Van~der Sande, G.}
\newblock \bibinfo{title}{A poor man’s coherent ising machine based on
  opto-electronic feedback systems for solving optimization problems}.
\newblock \emph{\bibinfo{journal}{Nature communications}}
  \textbf{\bibinfo{volume}{10}}, \bibinfo{pages}{3538} (\bibinfo{year}{2019}).

\bibitem{hamerly2019experimental}
\bibinfo{author}{Hamerly, R.}, \bibinfo{author}{Inagaki, T.},
  \bibinfo{author}{McMahon, P.~L.}, \bibinfo{author}{Venturelli, D.},
  \bibinfo{author}{Marandi, A.}, \bibinfo{author}{Onodera, T.},
  \bibinfo{author}{Ng, E.}, \bibinfo{author}{Langrock, C.},
  \bibinfo{author}{Inaba, K.}, \bibinfo{author}{Honjo, T.} \emph{et~al.}
\newblock \bibinfo{title}{Experimental investigation of performance differences
  between coherent ising machines and a quantum annealer}.
\newblock \emph{\bibinfo{journal}{Science advances}}
  \textbf{\bibinfo{volume}{5}}, \bibinfo{pages}{eaau0823}
  (\bibinfo{year}{2019}).

\bibitem{albash2018demonstration}
\bibinfo{author}{Albash, T.} \& \bibinfo{author}{Lidar, D.~A.}
\newblock \bibinfo{title}{Demonstration of a scaling advantage for a quantum
  annealer over simulated annealing}.
\newblock \emph{\bibinfo{journal}{Physical Review X}}
  \textbf{\bibinfo{volume}{8}}, \bibinfo{pages}{031016} (\bibinfo{year}{2018}).

\bibitem{denchev2016computational}
\bibinfo{author}{Denchev, V.~S.}, \bibinfo{author}{Boixo, S.},
  \bibinfo{author}{Isakov, S.~V.}, \bibinfo{author}{Ding, N.},
  \bibinfo{author}{Babbush, R.}, \bibinfo{author}{Smelyanskiy, V.},
  \bibinfo{author}{Martinis, J.} \& \bibinfo{author}{Neven, H.}
\newblock \bibinfo{title}{What is the computational value of finite-range
  tunneling?}
\newblock \emph{\bibinfo{journal}{Physical Review X}}
  \textbf{\bibinfo{volume}{6}}, \bibinfo{pages}{031015} (\bibinfo{year}{2016}).

\bibitem{boixo2016computational}
\bibinfo{author}{Boixo, S.}, \bibinfo{author}{Smelyanskiy, V.~N.},
  \bibinfo{author}{Shabani, A.}, \bibinfo{author}{Isakov, S.~V.},
  \bibinfo{author}{Dykman, M.}, \bibinfo{author}{Denchev, V.~S.},
  \bibinfo{author}{Amin, M.~H.}, \bibinfo{author}{Smirnov, A.~Y.},
  \bibinfo{author}{Mohseni, M.} \& \bibinfo{author}{Neven, H.}
\newblock \bibinfo{title}{Computational multiqubit tunnelling in programmable
  quantum annealers}.
\newblock \emph{\bibinfo{journal}{Nature communications}}
  \textbf{\bibinfo{volume}{7}}, \bibinfo{pages}{10327} (\bibinfo{year}{2016}).

\bibitem{sebastian2020memory}
\bibinfo{author}{Sebastian, A.}, \bibinfo{author}{Le~Gallo, M.},
  \bibinfo{author}{Khaddam-Aljameh, R.} \& \bibinfo{author}{Eleftheriou, E.}
\newblock \bibinfo{title}{Memory devices and applications for in-memory
  computing}.
\newblock \emph{\bibinfo{journal}{Nature nanotechnology}}
  \textbf{\bibinfo{volume}{15}}, \bibinfo{pages}{529--544}
  (\bibinfo{year}{2020}).

\bibitem{schroeder2022fundamentals}
\bibinfo{author}{Schroeder, U.}, \bibinfo{author}{Park, M.~H.},
  \bibinfo{author}{Mikolajick, T.} \& \bibinfo{author}{Hwang, C.~S.}
\newblock \bibinfo{title}{The fundamentals and applications of ferroelectric
  hfo2}.
\newblock \emph{\bibinfo{journal}{Nature Reviews Materials}}
  \textbf{\bibinfo{volume}{7}}, \bibinfo{pages}{653--669}
  (\bibinfo{year}{2022}).

\bibitem{salahuddin2018era}
\bibinfo{author}{Salahuddin, S.}, \bibinfo{author}{Ni, K.} \&
  \bibinfo{author}{Datta, S.}
\newblock \bibinfo{title}{The era of hyper-scaling in electronics}.
\newblock \emph{\bibinfo{journal}{Nature Electronics}}
  \textbf{\bibinfo{volume}{1}}, \bibinfo{pages}{442--450}
  (\bibinfo{year}{2018}).

\bibitem{trentzsch201628nm}
\bibinfo{author}{Trentzsch, M.}, \bibinfo{author}{Flachowsky, S.},
  \bibinfo{author}{Richter, R.}, \bibinfo{author}{Paul, J.},
  \bibinfo{author}{Reimer, B.}, \bibinfo{author}{Utess, D.},
  \bibinfo{author}{Jansen, S.}, \bibinfo{author}{Mulaosmanovic, H.},
  \bibinfo{author}{M{\"u}ller, S.}, \bibinfo{author}{Slesazeck, S.},
  \bibinfo{author}{Ocker, J.}, \bibinfo{author}{Noack, M.},
  \bibinfo{author}{Muller, J.}, \bibinfo{author}{Polakowski, P.},
  \bibinfo{author}{Schreiter, J.}, \bibinfo{author}{Beyer, S.},
  \bibinfo{author}{Mikolajick, T.} \& \bibinfo{author}{Rice, B.}
\newblock \bibinfo{title}{A 28nm hkmg super low power embedded nvm technology
  based on ferroelectric fets}.
\newblock In \emph{\bibinfo{booktitle}{2016 IEEE International Electron Devices
  Meeting (IEDM)}}, \bibinfo{pages}{11--5} (\bibinfo{organization}{IEEE},
  \bibinfo{year}{2016}).

\bibitem{taoka2021simulated}
\bibinfo{author}{Taoka, K.}, \bibinfo{author}{Misawa, N.},
  \bibinfo{author}{Koshino, S.}, \bibinfo{author}{Matsui, C.} \&
  \bibinfo{author}{Takeuchi, K.}
\newblock \bibinfo{title}{Simulated annealing algorithm \& reram device
  co-optimization for computation-in-memory}.
\newblock In \emph{\bibinfo{booktitle}{2021 IEEE International Memory Workshop
  (IMW)}}, \bibinfo{pages}{1--4} (\bibinfo{organization}{IEEE},
  \bibinfo{year}{2021}).

\bibitem{misawa2022domain}
\bibinfo{author}{Misawa, N.}, \bibinfo{author}{Taoka, K.},
  \bibinfo{author}{Matsui, C.} \& \bibinfo{author}{Takeuchi, K.}
\newblock \bibinfo{title}{Domain specific reram computation-in-memory design
  considering bit precision and memory errors for simulated annealing}.
\newblock In \emph{\bibinfo{booktitle}{2022 IEEE International Symposium on
  Circuits and Systems (ISCAS)}}, \bibinfo{pages}{3289--3293}
  (\bibinfo{organization}{IEEE}, \bibinfo{year}{2022}).

\bibitem{beyer2020fefet}
\bibinfo{author}{Beyer, S.}, \bibinfo{author}{D{\"u}nkel, S.},
  \bibinfo{author}{Trentzsch, M.}, \bibinfo{author}{M{\"u}ller, J.},
  \bibinfo{author}{Hellmich, A.}, \bibinfo{author}{Utess, D.},
  \bibinfo{author}{Paul, J.}, \bibinfo{author}{Kleimaier, D.},
  \bibinfo{author}{Pellerin, J.}, \bibinfo{author}{M{\"u}ller, S.}
  \emph{et~al.}
\newblock \bibinfo{title}{Fefet: A versatile cmos compatible device with
  game-changing potential}.
\newblock In \emph{\bibinfo{booktitle}{2020 IEEE International Memory Workshop
  (IMW)}}, \bibinfo{pages}{1--4} (\bibinfo{organization}{IEEE},
  \bibinfo{year}{2020}).

\bibitem{soliman2020ultra}
\bibinfo{author}{Soliman, T.}, \bibinfo{author}{M{\"u}ller, F.},
  \bibinfo{author}{Kirchner, T.}, \bibinfo{author}{Hoffmann, T.},
  \bibinfo{author}{Ganem, H.}, \bibinfo{author}{Karimov, E.},
  \bibinfo{author}{Ali, T.}, \bibinfo{author}{Lederer, M.},
  \bibinfo{author}{Sudarshan, C.}, \bibinfo{author}{K{\"a}mpfe, T.},
  \bibinfo{author}{Guntoro, A.} \& \bibinfo{author}{Wehn, N.}
\newblock \bibinfo{title}{Ultra-low power flexible precision fefet based analog
  in-memory computing}.
\newblock In \emph{\bibinfo{booktitle}{2020 IEEE International Electron Devices
  Meeting (IEDM)}}, \bibinfo{pages}{29--2} (\bibinfo{organization}{IEEE},
  \bibinfo{year}{2020}).

\bibitem{saito2021analog}
\bibinfo{author}{Saito, D.}, \bibinfo{author}{Kobayashi, T.},
  \bibinfo{author}{Koga, H.}, \bibinfo{author}{Ronchi, N.},
  \bibinfo{author}{Banerjee, K.}, \bibinfo{author}{Shuto, Y.},
  \bibinfo{author}{Okuno, J.}, \bibinfo{author}{Konishi, K.},
  \bibinfo{author}{Di~Piazza, L.}, \bibinfo{author}{Mallik, A.},
  \bibinfo{author}{Van~Houdt, J.}, \bibinfo{author}{Tsukamoto, M.},
  \bibinfo{author}{Ohkuri, K.}, \bibinfo{author}{Umebayashi, T.} \&
  \bibinfo{author}{Ezaki, T.}
\newblock \bibinfo{title}{Analog in-memory computing in fefet-based 1t1r array
  for edge ai applications}.
\newblock In \emph{\bibinfo{booktitle}{2021 Symposium on VLSI Technology}},
  \bibinfo{pages}{1--2} (\bibinfo{organization}{IEEE}, \bibinfo{year}{2021}).

\bibitem{graphcolorcmu}
\bibinfo{title}{{CMU graph coloring dataset}}.
\newblock \bibinfo{note}{Https://mat.tepper.cmu.edu/COLOR/instances.html}.

\bibitem{maxcutstanford}
\bibinfo{title}{{Starford Max-Cut dataset}}.
\newblock \bibinfo{note}{Http://web.stanford.edu/~yyye/yyye/Gset/}.

\bibitem{hong2021memory}
\bibinfo{author}{Hong, M.-C.}, \bibinfo{author}{Cho, L.-C.},
  \bibinfo{author}{Lin, C.-S.}, \bibinfo{author}{Lin, Y.-H.},
  \bibinfo{author}{Chen, P.-A.}, \bibinfo{author}{Wang, I.-T.},
  \bibinfo{author}{Tzeng, P.-J.}, \bibinfo{author}{Sheu, S.-S.},
  \bibinfo{author}{Lo, W.-C.}, \bibinfo{author}{Wu, C.-I.} \&
  \bibinfo{author}{Hou, T.-H.}
\newblock \bibinfo{title}{In-memory annealing unit (imau): Energy-efficient
  (2000 tops/w) combinatorial optimizer for solving travelling salesman
  problem}.
\newblock In \emph{\bibinfo{booktitle}{2021 IEEE International Electron Devices
  Meeting (IEDM)}}, \bibinfo{pages}{21--3} (\bibinfo{organization}{IEEE},
  \bibinfo{year}{2021}).

\bibitem{cai2020power}
\bibinfo{author}{Cai, F.}, \bibinfo{author}{Kumar, S.},
  \bibinfo{author}{Van~Vaerenbergh, T.}, \bibinfo{author}{Sheng, X.},
  \bibinfo{author}{Liu, R.}, \bibinfo{author}{Li, C.}, \bibinfo{author}{Liu,
  Z.}, \bibinfo{author}{Foltin, M.}, \bibinfo{author}{Yu, S.},
  \bibinfo{author}{Xia, Q.}, \bibinfo{author}{Yang, J.~J.},
  \bibinfo{author}{Beausoleil, R.}, \bibinfo{author}{Lu, W.~D.} \&
  \bibinfo{author}{Strachan, J.~P.}
\newblock \bibinfo{title}{Power-efficient combinatorial optimization using
  intrinsic noise in memristor hopfield neural networks}.
\newblock \emph{\bibinfo{journal}{Nature Electronics}}
  \textbf{\bibinfo{volume}{3}}, \bibinfo{pages}{409--418}
  (\bibinfo{year}{2020}).

\bibitem{yang2020transiently}
\bibinfo{author}{Yang, K.}, \bibinfo{author}{Duan, Q.}, \bibinfo{author}{Wang,
  Y.}, \bibinfo{author}{Zhang, T.}, \bibinfo{author}{Yang, Y.} \&
  \bibinfo{author}{Huang, R.}
\newblock \bibinfo{title}{Transiently chaotic simulated annealing based on
  intrinsic nonlinearity of memristors for efficient solution of optimization
  problems}.
\newblock \emph{\bibinfo{journal}{Science advances}}
  \textbf{\bibinfo{volume}{6}}, \bibinfo{pages}{eaba9901}
  (\bibinfo{year}{2020}).

\bibitem{shin2018hardware}
\bibinfo{author}{Shin, J.~H.}, \bibinfo{author}{Jeong, Y.~J.},
  \bibinfo{author}{Zidan, M.~A.}, \bibinfo{author}{Wang, Q.} \&
  \bibinfo{author}{Lu, W.~D.}
\newblock \bibinfo{title}{Hardware acceleration of simulated annealing of spin
  glass by rram crossbar array}.
\newblock In \emph{\bibinfo{booktitle}{2018 IEEE International Electron Devices
  Meeting (IEDM)}}, \bibinfo{pages}{3--3} (\bibinfo{organization}{IEEE},
  \bibinfo{year}{2018}).

\bibitem{mahmoodi2019analog}
\bibinfo{author}{Mahmoodi, M.}, \bibinfo{author}{Kim, H.},
  \bibinfo{author}{Fahimi, Z.}, \bibinfo{author}{Nili, H.},
  \bibinfo{author}{Sedov, L.}, \bibinfo{author}{Polishchuk, V.} \&
  \bibinfo{author}{Strukov, D.}
\newblock \bibinfo{title}{An analog neuro-optimizer with adaptable annealing
  based on 64$\times$ 64 0t1r crossbar circuit}.
\newblock In \emph{\bibinfo{booktitle}{2019 IEEE International Electron Devices
  Meeting (IEDM)}}, \bibinfo{pages}{14--7} (\bibinfo{organization}{IEEE},
  \bibinfo{year}{2019}).

\end{thebibliography}

\bibliographystyle{Nature}

\section*{Author contributions}
X.Y., T.K., and K.N. proposed and supervised the project. M.G., A.V., F.M., N.L., T.K. designed the chip and performed the experimental verification of the proposed design. Y.Q., Z.S.  and C.Z. conducted SPICE simulations and verification. Z.Z., Z.J., Y. S., X.G. helped with data analysis. All authors contributed to write up of the manuscript.

\section*{Competing Interests}
\label{sec:conflict}
The authors declare no conflicts of interest.

\newpage
\renewcommand{\thefigure}{S\arabic{figure}}
\renewcommand{\thetable}{S\arabic{table}}

\onecolumn
\centering
\textbf{\Large Supplementary Information}
\setcounter{figure}{0}
\setcounter{table}{0}
\setcounter{page}{1}

\begin{flushleft} 
\section{\textbf{\large Conversions from COPs  to  QUBO formulation}}
\label{sec:conversion}
\subsection{Max-Cut}
\end{flushleft}

\justify Max-Cut is a well-known optimization problem in computer science and graph theory. 
Given an undirected graph $G(V,E)$ composing of a vertex set $V$ and an edge set $E$, the object is to partition its vertices into two sets such that the number of edges crossing the partition is maximized \cite{glover2018tutorial}.

To convert a Max-Cut problem into the Quadratic Unconstrained Binary Optimization (QUBO) formulation, we define a binary variable $x_i \in \{0,1\}$ for each vertex $i$ in the graph, such that $x_i = 1$ if vertex $i$ is assigned to one set and $x_i = 0$ if it is assigned to the other set. 
The Max-Cut objective function can then be expressed as
\begin{equation}
\vspace{-1ex}
\label{equ:max cut0}
    \max \sum_{(i,j)\in E}(x_i + x_j-2x_i x_j)
\end{equation}
where $E$ is the set of edges in the graph. The term $(x_i + x_j-x_i x_j)$ takes the value 1 when the adjacent vertices $i$ and $j$ are assigned to different sets, and 0 otherwise. 
Without loss of generality, for all the edges $(i,j)$ in the graph, the object function  can be rewritten as
\begin{equation}
\vspace{-1ex}
\label{equ:max cut1}
    \min \sum_{(i,j)\in E}(2x_i x_j - x_i - x_j)
\end{equation}
which is in form of 
\begin{equation}
\vspace{-1ex}
\label{equ:max cut2}
    \min y = x^TQx
\end{equation}

\subsection{Graph Coloring}
The graph coloring problem is described as follows. 
Given an undirected graph $G=(V,E)$, where $V$ is the set of vertices and $E$ is the set of edges, the object is to assign a color to each vertex such  that each pair of  adjacent vertices have different colors. 
That is, to find a desired vertex coloring function  of a graph $G$ $f:V \rightarrow C$ where $C$ is a set of colors, such that $f(u) \neq f(v)$ for every pair of adjacent vertices $u$ and $v$ in $G$. 
Specifically, the K-coloring problem  whose object is to find a vertex coloring of a graph using  $K$ colors, is a classic problem in graph theory and has been extensively studied for a variety  of applications including frequency assignment problems and printed circuit board design problems \cite{glover2018tutorial}.

To convert a K-graph coloring problem into the QUBO formulation, we define a binary variable $x_{ip} \in \{0,1\}$ for a node 
coloring, such that $x_{ip} = 1$ if node $i$ is assigned color $p$, and 0 otherwise. 
Since each node must be colored by one color, we have the constraint
\begin{equation}
\vspace{-1ex}
\label{equ:graph coloring0}
    \sum_{p=1}^{K}x_{ip}=1 \quad i=1,...,n
\end{equation}
where n denotes the number of nodes in the graph.
All the edges $(m,n)$ in the graph are required to connect different colors, leading to the  constraint
\begin{equation}
\vspace{-1ex}
\label{equ:graph coloring1}
    x_{mp}+x_{np}\leq 1 \quad p=1,...,K
\end{equation}
The corresponding quadratic penalty for the constraint Eq. \ref{equ:graph coloring1} is assumed as
\begin{equation}
\vspace{-1ex}
\label{equ:quadratic penalties}
     x_{mp}x_{np} \quad p=1,...,K
\end{equation}
The penalty is not added only when the colors of adjacent nodes are  different.
The object function of graph coloring problem can then be expressed as a QUBO formulation:
\begin{equation}
\vspace{-1ex}
\label{equ:graph coloring2}
    \min y = \sum_{i \in V} \sum_{p}^K (x_{ip}-1)^2 + \sum_{(m,n) \in E } \sum_{p}^K x_{mp}x_{np}
\end{equation}

\subsection{Prime Factorization Problem}
The prime factorization problem (PFP) aims to find the prime factors of a large integer. 
Fig. \ref{fig:prime35} conceptually illustrates an example of converting to QUBO formulation.
Suppose we have an integer $N=P\times Q$ that needs to be factored, $P$ and $Q$ are firstly rewritten as $P = (1p_kp_{k-1}...p_11)_2$  and $Q = (1q_lq_{l-1}...q_11)_2$, respectively. 
The multiplication of $P\times Q$  is expanded as bit-wise partial products listed in the multiplication table. 
Then the bit-wise partial products of the two factors are grouped as integer blocks and sum to the corresponding integer blocks of the given integer $N$, forming the integer block equations. 
By introducing auxiliary variables, the objective function derived from the squared error of block equations is reduced to  QUBO formulation.

\begin{figurehere}
	\centering
	\includegraphics [width=0.8\linewidth]{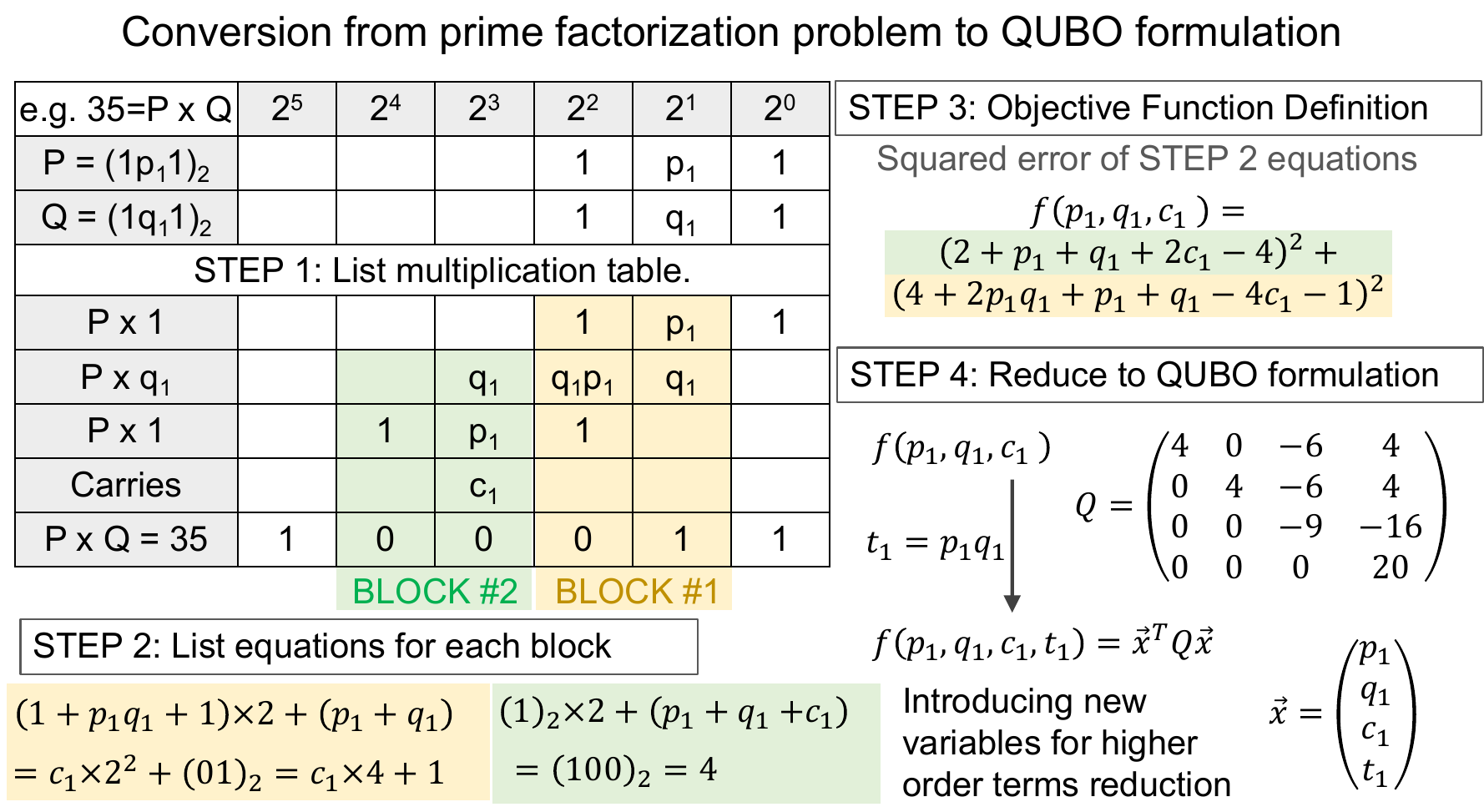}
	\caption{\textit{Conceptual conversion from prime factorization to  $\Vec{x}\ ^TQ\Vec{x}$ QUBO formulation by leveraging the block-wise binary multiplication table,  objective function formulation and higher order term reduction methods.}}
	\label{fig:prime35}
\end{figurehere}

\newpage
\begin{flushleft} 
\section{\textbf{\large FeFET CiM Chip Integration \& Testing}}
\label{sec:supp_chip}
\end{flushleft}

\justify The electrical characterization setup is shown in Fig. \ref{fig:meas_setup}\textbf{a}. The measurements are performed mainly with a PXIe measurement system from National Instruments (i.e., label 1 in Fig. \ref{fig:meas_setup}). A padring of 28 individual analog and digital pads connect the 1kb (32$\times$32) FeFET macro with a serial peripheral interface (SPI). The adapter board (label 2 in Fig.\ref{fig:meas_setup}) connects to the specific pads on the wafer in a wafer probe system (label 3 in Fig. \ref{fig:meas_setup}) via a probe-card (label 4 in Fig. \ref{fig:meas_setup}), see zoomed in for Fig. \ref{fig:meas_setup}\textbf{b}/\textbf{c}. 

A set of separate NI PXIe-4143 source measure units (SMU) and an NI PXIe-6570 pattern generator. Notably, the output pins of the latter device can be utilized as a Pin Parametric Measurement Unit (PPMU). Through this arrangement, the requisite supply, bias voltages, and digital signals are generated. Additionally, the pattern generator plays an instrumental role in forming the scan chain for the appropriate addressing of wordlines and sourceline/drainline.

For the task of loading vectors $x$ and $y$, an external clock, operating at an upper limit of 50 MHz, is employed. The wordline voltage designated for reading is sourced from the PPMU, covering both logical states '0' and '1'. In the context of this experiment, these states correspond to 0V and 1.2V, respectively, during the read phase. For this experimental demonstration, we posit a ternary precision for the $Q$ matrix. This matrix is depicted through the decomposition of ternary weights into a 2FeFET cell. This arrangement obviates the need for an ADC and shift-add operations. Consequently, all binary operations can be aggregated across the entire matrix within a singular cell, thereby streamlining and accelerating the overarching operation. The macro is equipped to support both single-bit reading and parallel decoders, ensuring maximal flexibility in wordline/bitline combinations. The matrix also features adaptable bitline multiplexers, facilitating the selection of the active bitlines. In culmination, the current across all active bitlines $y$ in the 1kb 1FeFET-1R array, and by extension the $xQy$ outcome, is ascertained in a singular read cycle with a designated bitline voltage of 0.1V.

The matrix parameters of $Q$ are programmed word by word into the FeFET array following a V\textsubscript{W}/3 inhibit scheme (V\textsubscript{W} is the write voltage). Specifically, we apply a pulse of 3.4V for the selected wordlines, 0.8V to unselected wordlines as well as 0V to selected and 1.8V to non-selected sourcelines with a duration of 1ms.

\begin{figurehere}
    
    \centering
    \includegraphics[width=0.8\columnwidth]{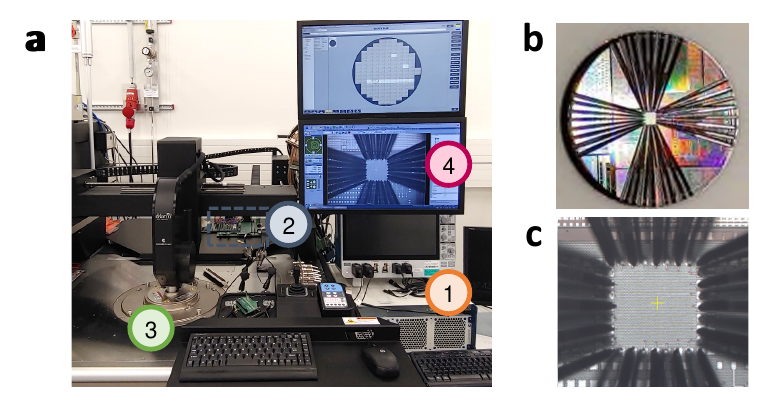}
    \caption{\textbf{Measurement setup for FeFET characterization.} A PXI System provides Source Measurement Units (SMU) and Pin Parametric Measurement Units (PPMU). PPMUs are used to configure the Switch Matrix for routing the source signals to the respective contact needles. Test structures are available on 300 mm wafers and connected to the measurement setup on a semi-automatic probe station using a probe card.}
    \label{fig:meas_setup}
\end{figurehere}

\newpage
\begin{flushleft} 
\section{\textbf{\large Lossless QUBO compression}}
\label{sec:supp_compression}
\end{flushleft}

\begin{figurehere}
	\centering
	\includegraphics [width=1\linewidth]{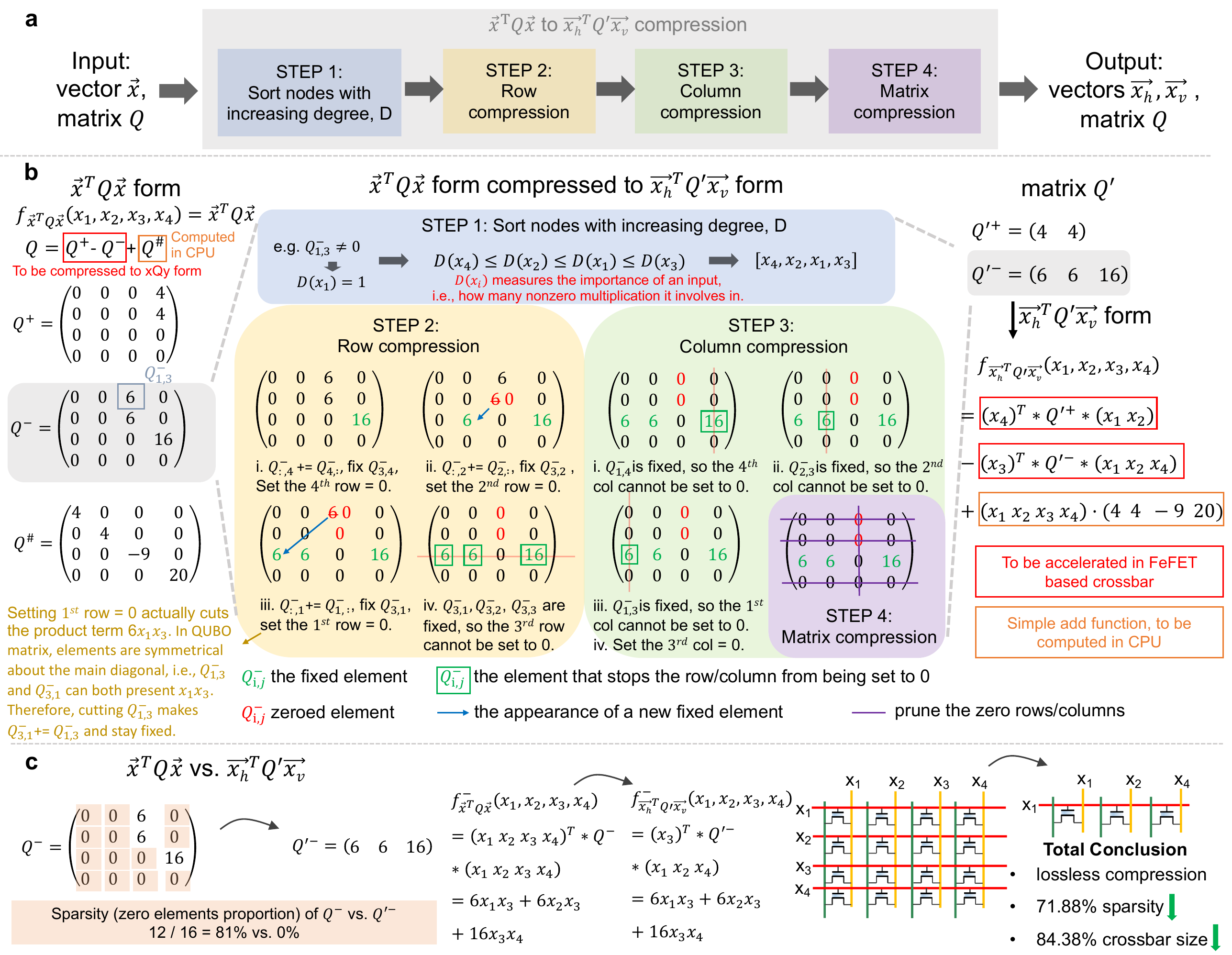}
	\caption{\textit{\textbf{a}, Flow of QUBO compression from $\Vec{x}\ ^TQ\Vec{x}$ to $\Vec{x_h}\ ^TQ'\Vec{x_v}$ QUBO formulation. \textbf{b}, Compression example of the PFP 35=5$\times$7 QUBO matrix; \textbf{c}, The significant savings of the lossless QUBO compression in \textbf{b}, respectively.}}
	\label{fig:compression example}
\end{figurehere}

\justify Fig. \ref{fig:compression example}a and b illustrate the QUBO compression flow and a detailed example of the PFP converted QUBO compression, respectively.
Following the compression steps described in the main text, the QUBO matrixes converted from the PFP 35=5$\times$7 are compressed as below (take $Q^-$ as the example):

\textbf{STEP 1}: All the input variables  $\Vec{x}$  are sorted in a connectivity increasing order, i.e., CIA=$[x4, x2, x1, x3]$.

\textbf{STEP 2}: Row compression of Q matrix is performed in the order of CIA.
For  variable $x_4$ selected from CIA,  its  
corresponding $4^{th}$ row contains no 
\textit{fixed} element, thus this row can be compressed, denoted as compressed row. 
All the elements $Q^-_{4,j}, j\in [1,4]$ within the row are added to their diagonal elements, i.e., $Q^-_{j,4} += Q^-_{4,j}, j\in [1,4]$.
The diagonal element $Q^-_{3,4}$ is nonzero, thus \textit{fixed} to ensure the presence of the associated product term, i.e., $16x_3x_4$ in QUBO formulation.
For variable $x_2$,  the corresponding $2^{nd}$ row contains no \textit{fixed} element, this row can be compressed, denoted as compressed row.
The elements $Q^-_{2,j}, j\in [1,4]$ are added to their diagonal element $Q^-_{j,2}$, set to 0, and then their diagonal nonzero element $Q^-_{3,2}$ is \textit{fixed} to keep  $6x_3x_2$ in QUBO formulation.
For variable $x_1$, its corresponding row contains no \textit{fixed} element, thus can be compressed.
The elements $Q^-_{1,j}, j\in [1,4]$ are added to the diagonal elements $Q^-_{j,1}$, set to 0 and the diagonal nonzero element $Q^-_{3,1}$ is \textit{fixed} to keep $6x_3x_1$ in QUBO formulation.
For the last variable $x_3$, its corresponding row contains \textit{fixed} elements $Q^-_{3,1}$, $Q^-_{3,2}$ and $Q^-_{3,4}$, thus cannot be compressed.



\textbf{STEP 3}: 
Column compression  follows the similar operation to the row compression in Step 2.
Following the CIA, the $4^{th}$, $2^{nd}$, $1^{st}$ columns contain  \textit{fixed} elements $Q^-_{3,4}$, $Q^-_{3,2}$, $Q^-_{3,1}$, respectively, thus cannot be compressed.
The $3^{rd}$ column contains no \textit{fixed} elements, thus denoted as compressed column.
The elements within the column are zero, and their diagonal elements are zero, thus the $3^{rd}$ column can be set to 0.


\textbf{STEP 4}: The  compressed rows, i.e., the $1^{st}$, $2^{nd}$, $4^{th}$ rows and the compressed column, i.e., the $3^{rd}$ column of the QUBO matrix along with their respective variables  in the  input vectors, i.e., $x_1$, $x_2$, $x_4$ in $\Vec{x_h}$ and $x_3$ in $\Vec{x_v}$, are removed, forming the compressed QUBO formulation, i.e., $x_3^TQ^{'-}(x_1, x_2, x_4)$, $Q^{'-}=(6, 6, 16)$.

Fig. \ref{fig:compression example}c shows the compressed QUBO formulation and its crossbar size used to map the matrix. The binary variable vectors $\Vec{x_h}$/$\Vec{x_v}$ are applied to the WLs/SLs of the FeFET crossbar array to implement iterative vector-matrix-vector multiplications. 
In this PFP example, two QUBO matrixes ($Q^+$ and $Q^-$) are compressed, achieving 71.88\% sparsity reduction and 84.38\% chip area saving compared to the original QUBO implementation. 


\newpage
\begin{flushleft} 
\section{\textbf{\large Example: MESA Analysis over Max-Cut problem}}
\label{sec:supp_MESA}
\end{flushleft}

\begin{figurehere}
	\centering
	\includegraphics [width=0.5\linewidth]{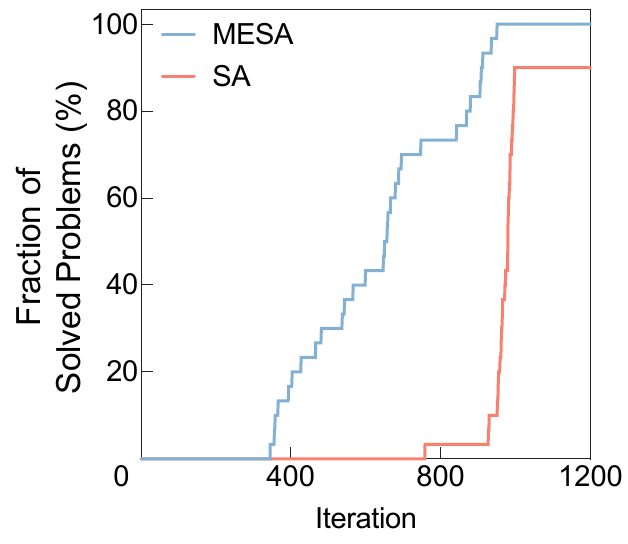}
	\caption{\textit{
    The fraction of solved problems and time-to-solution of MESA compared to conventional SA in solving  Max-Cut problems.
    }}
	\label{fig: supp MESA}
\end{figurehere}

\justify Fig. \ref{fig: supp MESA} depicts comparison between MESA and conventional SA in solving  Max-Cut problems.
Results show that 
the fraction of  solved problems by MESA gradually grows with iteration count, and eventually reaches  100\%. 
In contrast, the fraction of solved problems by conventional SA remains 0 until nearly 1,000 iterations, and ultimately reaches 90\% with more iteration count.

\newpage
\begin{flushleft} 
\section{\textbf{\large Example: prime factorization problem (PFP) of 323}}
\label{sec:supp_PFP}
\end{flushleft}

\justify Fig. \ref{fig:SA 323}a shows the MESA process in searching for the solution of factoring 323 within one SA epoch, where 
the number of input variables is 10, and the QUBO coefficient precision is set as 5-bit. 
The corresponding  energy evolution of the converted QUBO formulation is shown in Fig. \ref{fig:SA 323}b.
Fig. \ref{fig:SA 323}c shows the success rates of solving the PFP of 323, which is defined  as the probability of finding the optimal factors.
The results suggest that insufficient weight precision stored in the compute-in-memory crossbar may lead to imprecise solutions, as the  energy landscape with quantized QUBO formulation deviates significantly from the real energy landscape. 
High success rate can be achieved with 5-bit precision configuration.  
Fig. \ref{fig:SA 323}d shows that most of the minimal attractors found by our proposed framework with MES are validated to be optimal/correct solutions. 

\begin{figurehere}
	\centering
	\includegraphics [width=0.8\linewidth]{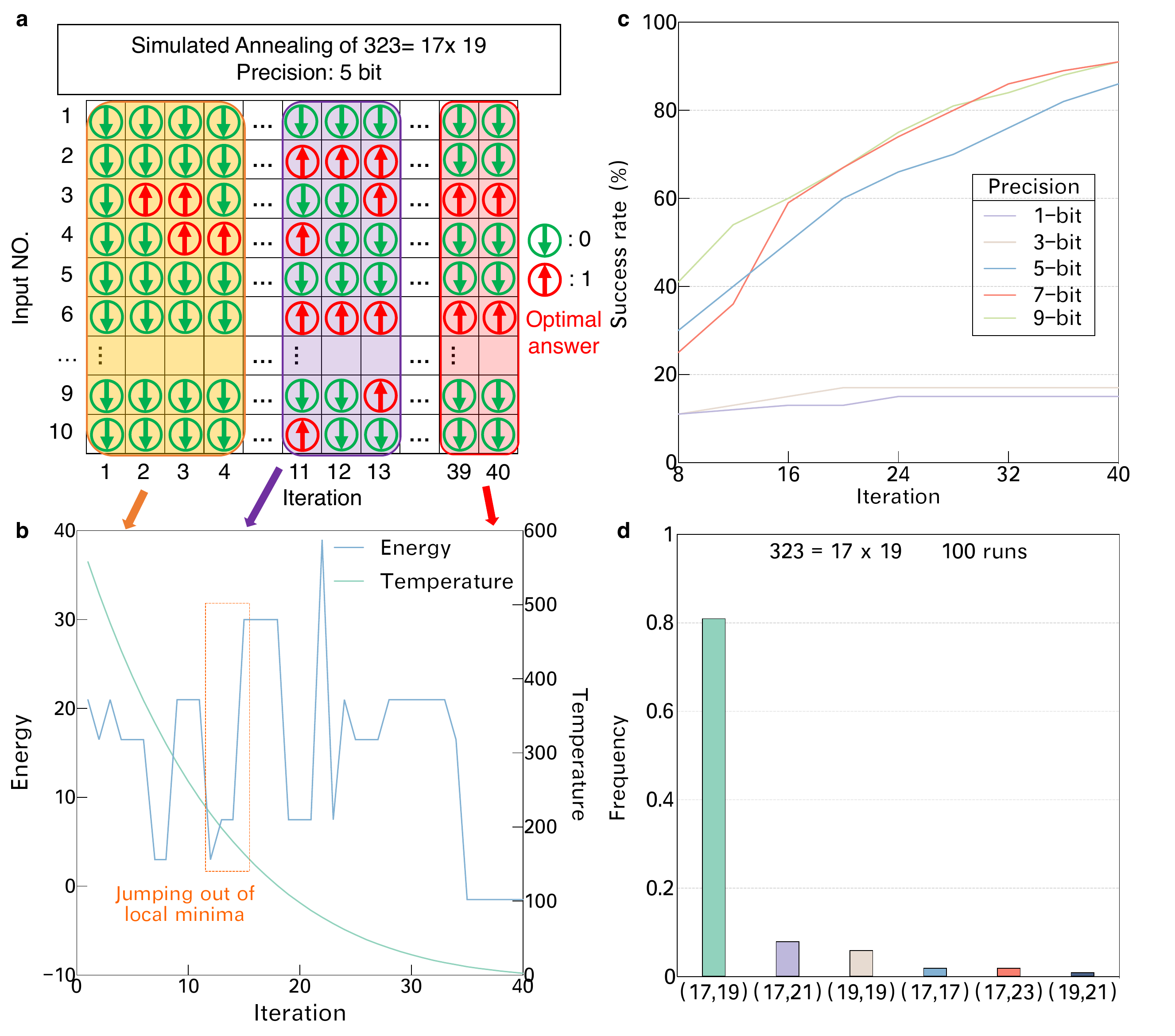}
	\caption{\textit{\textbf{a}, Exemplar MESA process and \textbf{b}, the corresponding  energy evolution process of PFP solving.
	\textbf{c}, The success rates of factoring 323 with different weight precisions. \textbf{d}, Attractor occurrences for factoring 323. }}
	\label{fig:SA 323}
\end{figurehere}

To achieve a higher success rate for solving PFPs, increasing the QUBO coefficient precision is an option, as illustrated in Fig. \ref{fig: 323 precision}.
That said, higher precision requires higher analog-digital-converter (ADC) resolution to accurately compute the QUBO formulation. Therefore, higher success rate comes at the cost of extra energy and hardware overheads.

\begin{figurehere}
	\centering
	\includegraphics [width=1\linewidth]{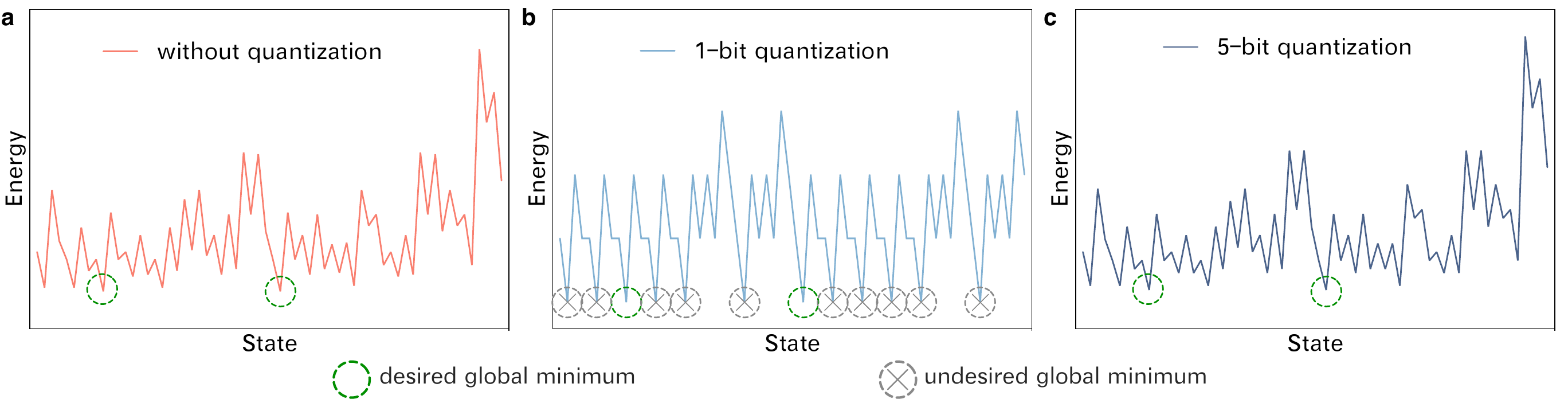}
	\caption{\textit{To \textbf{a}, The ideal energy landscape of QUBO formulation when solving 323 = 17$\times$19. \textbf{b}, Low bit quantization of QUBO coefficient results in numerous erroneous global minima.\textbf{c}, The landscape under high bit quantization exhibits no erroneous global minima, thus enhancing the success rate, as illustrated in Fig. \ref{fig:SA 323}c.}}
	\label{fig: 323 precision}
\end{figurehere}

\newpage
\begin{flushleft} 
\section{\textbf{\large Experimental demonstration of graph coloring problem}}
\label{sec:supp_graph}
\end{flushleft}

\justify Fig. \ref{fig: die all} demonstrates the QUBO energy functions measured on our fabricated FeFET crossbar prototypes for solving the exemplar graph coloring problem depicted in Fig. \ref{fig:result}. 
As can be seen from the energy trajectories, all three prototypes are capable of  finding the optimal solution of the problem within 100 iterations, thus suggesting the robustness of both our prototypes and the proposed hardware-algorithm co-design framework. 
Moreover, the intra-prototype variation of energy curves is negligible, highlighting the performance consistency of our fabricated hardware.

\begin{figurehere}
	\centering
	\includegraphics [width=1\linewidth]{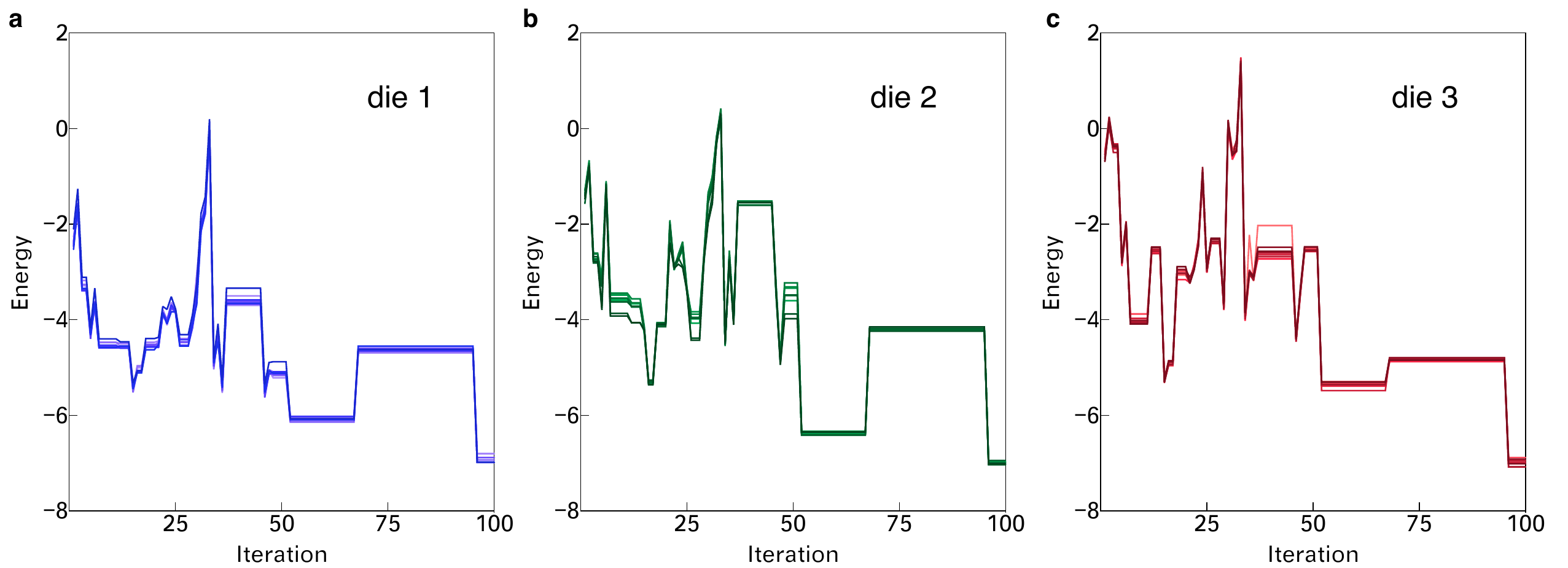}
	\caption{\textit{Experimental results for solving the graph coloring problem, as shown in Fig. \ref{fig:result}f.
	The energy curves for three different dies \textbf{a}/\textbf{b}/\textbf{c}, with each die running the problem-solving process 9 times.}}
	\label{fig: die all}
\end{figurehere}

\justify Fig. \ref{fig: graphcolor abc}
showcases the graph coloring configurations during annealing, highlighted at different iteration steps shown in Fig.\ref{fig:result}\textit{i}, i.e., the beginning (A), midpoint (B), and end (C) of the evolution process.

\begin{figurehere}
	\centering
	\includegraphics [width=1\linewidth]{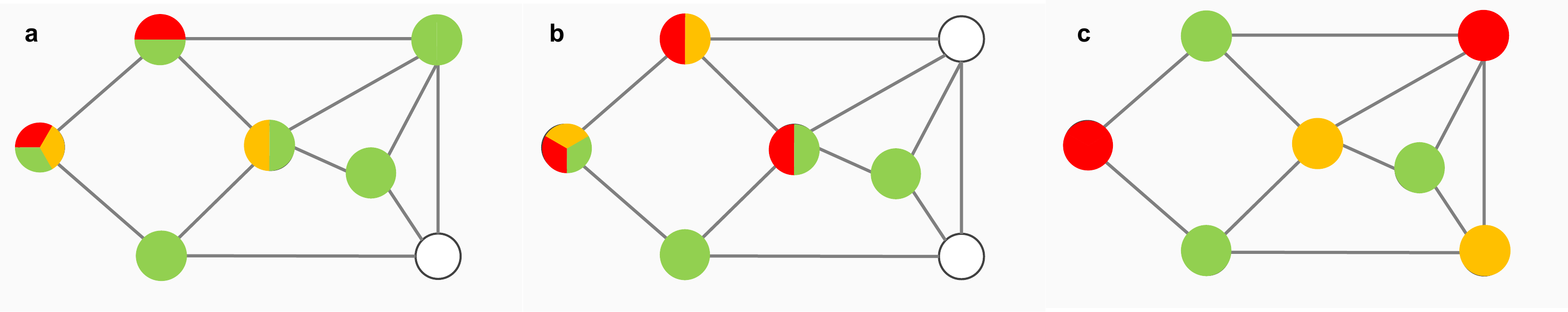}
	\caption{\textit{Graph coloring configurations at annealing stages (\textbf{a}, \textbf{b}, \textbf{c}) during the evolution process. 
 White means no color assignment. Nodes with one, two, three colors means that one, two, or three colors have been assigned to the nodes. Note that more than one color does not meet the constraint and no longer exists at the final converged solution.}}
	\label{fig: graphcolor abc}
\end{figurehere}

\end{document}